\newif\ifpreprint
\preprinttrue

\ifpreprint
    \documentclass[manuscript,screen,nonacm]{acmart}
\else
    \documentclass[manuscript,review,anonymous]{acmart}
\fi
\AtBeginDocument{%
  }

\usepackage{xcolor}
\usepackage{soul}

\usepackage{amssymb}
\usepackage{booktabs}
\usepackage{colortbl}
\usepackage{graphicx}
\usepackage{multirow}
\usepackage{tabularx}
\usepackage{array}
\usepackage{makecell}
\usepackage{subcaption}

\usepackage[table,xcdraw]{xcolor}

\definecolor{editcol}{HTML}{FF0000}
\newcommand{\majorr}[1]{{\color{black}#1}}

\newcommand{\factorsymbol}[1]{%
  \raisebox{0.1ex}{\scriptsize\ensuremath{#1}}%
}

\newcommand{\levelsymbol}[1]{%
  \raisebox{0.3ex}{%
    \scalebox{0.60}{\ensuremath{#1}}%
  }%
}

\newcommand{\factorstyle}[3]{%
  \textcolor{#1}{\factorsymbol{#2}}%
  \setulcolor{#1}%
  \setul{0.35ex}{0.85pt}%
  \nobreakspace\texttt{\ul{#3}}%
}

\newcommand{\levelstyle}[3]{%
  \textcolor{#1}{\levelsymbol{#2}}%
  \nobreakspace
  \begingroup
  \texttt{#3}%
  \endgroup
}

\definecolor{AIPresenceColor}{HTML}{E69F00}
\definecolor{TaskTypeColor}{HTML}{0072B2}
\definecolor{RoleColor}{HTML}{D55E00}
\definecolor{SpecColor}{HTML}{009E73}
\definecolor{VisColor}{HTML}{CC79A7}

\DeclareRobustCommand{\aipresence}[1][AI Presence]{%
  \factorstyle{AIPresenceColor}{\bigstar}{#1}%
}
\DeclareRobustCommand{\aipresencelevel}[1]{%
  \levelstyle{AIPresenceColor}{\bigstar}{#1}%
}

\DeclareRobustCommand{\tasktype}[1][Task Type]{%
  \factorstyle{TaskTypeColor}{\bullet}{#1}%
}
\DeclareRobustCommand{\tasktypelevel}[1]{%
  \levelstyle{TaskTypeColor}{\bullet}{#1}%
}

\DeclareRobustCommand{\role}[1][Role]{%
  \factorstyle{RoleColor}{\blacklozenge}{#1}%
}
\DeclareRobustCommand{\rolelevel}[1]{%
  \levelstyle{RoleColor}{\blacklozenge}{#1}%
}

\DeclareRobustCommand{\specialization}[1][Specialization]{%
  \factorstyle{SpecColor}{\blacktriangle}{#1}%
}
\DeclareRobustCommand{\specializationlevel}[1]{%
  \levelstyle{SpecColor}{\blacktriangle}{#1}%
}

\DeclareRobustCommand{\visibility}[1][Visibility]{%
  \factorstyle{VisColor}{\blacksquare}{#1}%
}
\DeclareRobustCommand{\visibilitylevel}[1]{%
  \levelstyle{VisColor}{\blacksquare}{#1}%
}

\DeclareRobustCommand{\aienabled}[1][AI-enabled]{%
  \aipresencelevel{#1}%
}
\DeclareRobustCommand{\nonai}[1][Non-AI]{%
  \aipresencelevel{#1}%
}

\DeclareRobustCommand{\assurance}[1][Assurance]{\tasktypelevel{#1}}
\DeclareRobustCommand{\brainstorming}[1][Brainstorming]{\tasktypelevel{#1}}
\DeclareRobustCommand{\proofing}[1][Proofing]{\tasktypelevel{#1}}
\DeclareRobustCommand{\search}[1][Search]{\tasktypelevel{#1}}

\DeclareRobustCommand{\assisting}[1][Assisting]{\rolelevel{#1}}
\DeclareRobustCommand{\delegating}[1][Delegating]{\rolelevel{#1}}

\DeclareRobustCommand{\openassist}[1][OpenAssist]{\specializationlevel{#1}}
\DeclareRobustCommand{\corpassist}[1][CorpAssist]{\specializationlevel{#1}}

\DeclareRobustCommand{\personal}[1][Personal]{\visibilitylevel{#1}}
\DeclareRobustCommand{\shared}[1][Shared]{\visibilitylevel{#1}}

\newcommand{\dv}[1]{\textsc{#1}}
\newcommand{\dvhs}[1][Help-Seeking]{\dv{#1}}
\newcommand{\dvcc}[1][Colleague Connection]{\dv{#1}}

\definecolor{elecblue}{rgb}{0.0, 0.5, 1.0}
\definecolor{lightblue}{rgb}{0.53, 0.81, 0.98}
\usepackage[textsize=scriptsize]{todonotes}

\definecolor{editCol}{RGB}{0,0,0}
\definecolor{lightline}{RGB}{210,210,210}

\ifpreprint
    \setcopyright{none}
\else
\setcopyright{acmlicensed}
\copyrightyear{2018}
\acmYear{2018}
\acmDOI{XXXXXXX.XXXXXXX}
\acmConference[Conference acronym 'XX]{Make sure to enter the correct
  conference title from your rights confirmation email}{June 03--05,
  2018}{Woodstock, NY}
\acmISBN{978-1-4503-XXXX-X/2018/06}
\fi

\begin{document}

\title[AI-Induced Social Erosion]{AI-Induced Social Erosion: Using LLMs for Work Tasks Reduces Help-Seeking Intentions Towards Colleagues}

\author{Kevin Chow}
\email{k.chow@nyu.edu} 
\orcid{0009-0008-5455-3328}
\affiliation{%
  \institution{New York University}
  \department{Technology Management and Innovation, Tandon School of Engineering}
  \city{Brooklyn}
  \state{New York}
  \country{USA}}

\author{Sophia Deng}
\email{sd5718@nyu.edu} 
\orcid{0009-0003-2945-7389}
\affiliation{%
  \institution{New York University}
  \department{Technology Management and Innovation, Tandon School of Engineering}
  \city{Brooklyn}
  \state{New York}
  \country{USA}}

\author{Bowen He}
\email{bh3044@nyu.edu} 
\orcid{0009-0006-5359-6903}
\affiliation{%
  \institution{New York University}
  \department{Technology Management and Innovation, Tandon School of Engineering}
  \city{Brooklyn}
  \state{New York}
  \country{USA}}

\author{Hanyi Min}
\email{hanyimin@illinois.edu} 
\orcid{0000-0002-0095-8513}
\affiliation{%
  \institution{University of Illinois Urbana-Champaign}
  \department{School of Labor and Employment Relations}
  \city{Urbana}
  \state{Illinois}
  \country{USA}}

\author{Qiaosi Wang}
\email{qiaosiw@andrew.cmu.edu} 
\orcid{0000-0002-5296-5440}
\affiliation{%
  \institution{Carnegie Mellon University}
  \department{Human-Computer Interaction Institute}
  \city{Pittsburgh}
  \state{Pennsylvania}
  \country{USA}}

\author{Koustuv Saha}
\email{ksaha2@illinois.edu}
\orcid{0000-0002-8872-2934}
\affiliation{%
  \institution{University of Illinois Urbana-Champaign}
  \department{Siebel School of Computing and Data Science}
  \city{Urbana}
  \state{Illinois}
  \country{USA}}

\author{Vedant Das Swain}
\email{vd2538@nyu.edu}
\orcid{0000-0001-6871-3523}
\affiliation{%
  \institution{New York University}
  \department{Technology Management and Innovation, Tandon School of Engineering}
  \city{Brooklyn}
  \state{New York}
  \country{USA}}

\renewcommand{\shortauthors}{Chow et al.}

\begin{abstract}
Generative AI is becoming embedded in knowledge work as an intelligent collaborator to assist in completing tasks. However, this readily available assistance risks displacing key interactions through which colleague relationships are built and sustained at work. This paper leverages the concept of help-seeking behaviour to investigate whether and how AI usage shapes colleague relationships. We conducted an online experimental vignette study where 248 knowledge workers evaluated scenarios that systematically varied whether and how AI was used across different work tasks. Using linear mixed-effects models, we found that AI-enabled work reduced help-seeking intentions compared to non-AI work across tasks, and especially for brainstorming. Visible AI usage was associated with greater help-seeking intentions towards colleagues, and individual contributors' colleague relationships were more vulnerable than managers. Our findings identify opportunities for workplace AI tools, practices, and policies to preserve colleague relationships in increasingly AI-integrated workplaces.

\end{abstract}

\ifpreprint
    \begin{center}
        \small\textit{\textcolor{red}{
            Note: This is a preprint --- this version of the manuscript has not been peer reviewed.
        }}
    \end{center}
    \vspace{1.5em}
\fi





\maketitle

\section{Introduction}

In 1986, Kelley and Caplan \cite{kelley1993bell} sought to understand how Bell Labs, then one of the world's most innovative research laboratories, produced so many high-performing knowledge workers. Contrary to expectations that star performers possessed greater cognitive capabilities, they instead found that these workers proactively built networks of colleagues through which they exchanged ideas, sought help, shared expertise, and supported one another. Decades of subsequent research have reinforced the importance of such strong colleague relationships at work, linking them to better job performance \cite{chiaburuPeersMakePlace2008}, creativity \cite{boekhorstFunFriendsCreativity2021,reich2011interpersonal}, job satisfaction \cite{winsteadQualityFriendshipsWork1995,simonGoodCompanyMultistudy2010,alegreAntecedentsEmployeeJob2016}, and overall well-being \cite{craigExaminingRoleFriendship2019,charoensukmongkolRoleCoworkerSupervisor2016}.

Generative AI\footnote{In this paper, we use AI and large-language models (LLMs) interchangeably with \textit{generative AI}, reflecting common contemporary vernacular.} is becoming embedded in how the modern knowledge worker accomplishes their work tasks. As of 2026, Gallup reported that 30\% of U.S. employees use AI in their role a few times a week or more \cite{incIndicatorArtificialIntelligence}. While the potential of such tools in the workplace has been recognized by many: they can help workers become more efficient \cite{noyExperimentalEvidenceProductivity2023,pengImpactAIDeveloper2023}, produce more \cite{brynjolfssonGenerativeAIWork2025,dellacquaNavigatingJaggedTechnological2023}, and access information or assistance on demand \cite{dellacquaCyberneticTeammateField2026}. 
At the same time that AI is increasingly being embedded into work, workplaces continue to face a ``loneliness epidemic'' \cite{figueiredoLonelinessIsolationEra2025,firozDesolatedMilieuExploring2020,murthy2017work}.  
This raises important questions about whether AI usage could have unintentional consequences related to workers' relationships with their colleagues, including both their sense of belonging and connection as well as their social capital \cite{nahapietSocialCapitalIntellectual1998}. 
By providing an always accessible and non-judgmental source of help, AI may reduce workers' reliance on their colleagues and displace regular social interactions by which relationships are built \cite{baygiGenerativeAISocial2026}. Although exchanging help between colleagues at work can be a slow, effortful, and vulnerable process \cite{leeSocialCostsSeeking2002,baygiGenerativeAISocial2026} these behaviours are characteristic of stronger, more trusting, and psychologically safe teams \cite{edmondsonPsychologicalSafetyLearning1999}. However, little empirical research has directly examined whether using AI, compared with completing work tasks without it, shapes relationships among colleagues.

In addition, AI usage is highly heterogeneous across individuals and organizations, with shared norms and best practices remaining underdeveloped \cite{balaynFrameworkCharacterizeReporting2026, bick2026rapid} --- meaning that the social effects of AI usage are unlikely to be uniform. For example, using LLMs to address routine questions may reduce unnecessary interruptions and mitigate collaborative overload \cite{crossCollaborativeOverload2016}, potentially preserving colleague relationships. Conversely, using AI in a manner that leads to passing on ``workslop'' --- poorly-vetted AI-generated output that requires significant additional effort from colleagues to interpret or correct --- may create frustration and undermine trust \cite{liebscherWorkslopExaminingPrevalence,limWorkslopWorkHow2026}. These examples suggest that AI's social consequences among colleagues may depend on how it is being used at work, yet it remains unclear which aspects of AI use might protect or strain colleague relationships.

Taken together, our overarching research question is: \textbf{How does knowledge workers' use of AI shape their working relationships with their colleagues?} We further examine two sub-questions: 

\begin{itemize}
    \item \textbf{RQ1}: How does the presence of AI-enabled work, compared to non-AI work, affect colleague relationships across different work tasks?
    \item \textbf{RQ2}: How does the way AI is being used for work tasks relate to the quality of colleague relationships?
\end{itemize}

To answer these questions, we conducted an experimental vignette study \cite{aguinisBestPracticeRecommendations2014} to systematically evaluate how different factors around AI usage influence colleague relationships. We operationalized colleague relationships through \textit{help-seeking} behaviour, inspired by the foundational Team-Member Exchange, a measure of worker relationship quality \cite{seersTeammemberExchangeQuality1989}. 
On a custom-designed online portal, participants evaluated a set of vignettes around AI usage for work, before answering survey questions on their subsequent help-seeking intentions towards their colleagues.  
We used linear mixed-effects models to analyze 2444 vignette evaluations from 248 knowledge worker participants, finding that AI-enabled work does lead to weaker help-seeking intentions and colleague connection, especially in brainstorming tasks. In AI-enabled work, using LLMs in a shared workspace that is visible to colleagues protects against reductions in help-seeking. 

We discuss how AI-enabled work might shift the cost-benefit calculus of help-seeking, the potential pitfalls of simply making all AI use visible across teams, and present several organizational opportunities to facilitate appropriate help-seeking even in an AI-embedded workspace. We make an \textit{empirical} contribution: experimentally demonstrating that AI-enabled work erodes behaviours that sustain colleague relationships (\textbf{RQ1}), and that these relational effects vary primarily by what task AI is being used for and the visibility of AI use (\textbf{RQ2}). 
Overall, our paper informs efforts to design and use AI at work so that, even as help-seeking becomes less frequent, it remains purposeful and is ultimately preserved where it matters most: to facilitate effective task support and forge meaningful relationships among colleagues.

\section{Background and Related Work}

We begin by reviewing help-seeking at work and describing how workers decide whether to seek help from their colleagues. Next, we examine how technologies, such as generative AI, have shaped help-seeking at work. However, with the limited and emerging research on AI and workplace help-seeking specifically, we round out our review by considering prior literature on how AI has shaped social connections more broadly, including outside of the workplace.

\subsection{What is Help-Seeking at Work?}

Social connections at work are critical to both organizational success and worker flourishing. Strong colleague relationships are associated with greater job performance, satisfaction, and personal well-being \cite{chiaburuPeersMakePlace2008,alegreAntecedentsEmployeeJob2016,craigExaminingRoleFriendship2019}. Such relationships also offer resources that buffer workplace stress and contribute to psychological safety \cite{colbertFlourishingWorkplaceRelationships2016,jollySocialSupportWork2021,edmondsonPsychologicalSafetyLearning1999}. In this study, we hone in on \textit{help-seeking} as one such activity through which colleague connections can be developed over time \cite{vanderrijtAskingHelpRelational2013,leeWhenGoingGets1997}. 

Help-seeking at work has been defined as an interpersonal, proactive, and discretionary process by which workers approach colleagues to help address and solve work-related problems \cite{bambergerEmployeeHelpseekingAntecedents2009, leeWhenGoingGets1997, chenHelpseekingWorkIntegrative2024,nadlerPersonalityHelpSeeking1997}.
It has been associated with a range of positive individual and organizational outcomes, including improved job performance \cite{gellerImpactHelpSeeking2012} and creativity \cite{muellerWhySeekingHelp2011, wangBoostingCreativityFunctional2022}, and is closely intertwined with workplace trust and supportive colleague relationships \cite{vanderrijtAskingHelpRelational2013, farmerContextualizedSelfHow2015}.
Furthermore, the exchange of help, information, and feedback is considered a key component of the aforementioned Team-Member Exchange (TMX) \cite{seersTeammemberExchangeQuality1989, seersTeamMemberExchangeTeam1995}, a foundational construct in the organizational behaviour literature that measures the quality of the working relationship between a worker and their teammates. However, seeking help, especially when done in an overly dependent and reliant manner \cite{nadler1997autonomous, bambergerEmployeeHelpseekingAntecedents2009}, can create demands on colleagues' time, attention, and energy, and even a sense of indebtedness \cite{hetrickHelpNotWanted2020, zhanFeelingGratitudeDepletion2021, greenberg1980theory}. These demands can reflect concerns around ``collaborative overload'' \cite{kimConsequencesCollaborativeOverload2023, crossCollaborativeOverload2016}, in which collaboration and helping demands lead to employee exhaustion and overwork. On the other hand, routine everyday exchanges of help are also an important means and social touchpoint by which workers can build and sustain colleague connections.

When deciding whether or not to seek help from their colleagues, workers face what has been coined as the \textit{help-seeker's dilemma} \cite{nadlerPersonalityHelpSeeking1997, cleavengerOvercomingHelpSeekersDilemma2015}: help-seekers must weigh the potential benefits of receiving help against the psychological and social costs of asking for it. Costs particularly centre around self-presentation \cite{goffman1978presentation}, as seeking help requires acknowledging a need or inability to independently resolve a problem \cite{leeWhenGoingGets1997, chenHelpseekingWorkIntegrative2024}. When doing so, workers may risk appearing less competent to their peers or supervisors, with these self-presentation concerns further complicated by status and power dynamics \cite{friedmanWhenDoesRespectful2018}. In an integrative review of workplace help-seeking, Chen et al.  \cite{chenHelpseekingWorkIntegrative2024} synthesized several antecedents that shape help-seeking decision-making: characteristics of the help-seeker, the target helper, the task itself, leadership and organizational culture, and broader environmental and temporal factors. Formal helping roles can signal a potential target helper's expertise and accessibility, which influences who workers approach for help \cite{hofmannSeekingHelpShadow2009, vanderrijtAskingHelpRelational2013}. In addition, Cleavenger et al. \cite{cleavengerHelpseekingTestingEffects2007} found that helping norms shaped help-seeking behaviours more strongly than task characteristics or individual traits such as self-esteem.  
However, much of the foundational research was conducted in face-to-face settings. In modern workplaces, technologies, including generative AI, are increasingly integrated into both how work is accomplished and how workers communicate and collaborate with one another, potentially shaping the costs and benefits workers navigate in making help-seeking decisions. 

\subsection{How Technology Has Shaped Help-Seeking at Work}

Workplace technologies have long shaped how employees access information, communicate, and collaborate with one another. Email, enterprise social media, and workplace messaging platforms have made colleagues easier to reach even across different work arrangements, including hybrid and remote work, potentially reducing some of the practical costs of help-seeking \cite{bambergerEmployeeHelpseekingAntecedents2009}. At the same time, technologies that provide easier access to broader information and organizational knowledge may also reduce the need to seek help from colleagues in the first place \cite{subramaniCapabilityDevelopmentJustinTime2021, oeldorf-hirschSearchAskRouting2014,ehrlichSearchingExpertise2008}. Yet, empirical evidence on technology's effects on workplace help-seeking remains limited, with one early study finding that greater anonymity encouraged employees to seek help more quickly \cite{cleavengerOvercomingHelpSeekersDilemma2015}, consistent with evidence from non-work contexts around lowering help-seeking and disclosure-related barriers, such as social media \cite{maEffectsAnonymityEphemerality2019} and mental health support systems \cite{choudhuryMentalHealthDiscourse2014} --- empirically supporting the impact of lowering social costs on facilitating help-seeking.

Generative AI makes the effects of technology on help-seeking especially salient as it may further alter both workers' for assistance and the costs of seeking it. 
LLMs can provide a highly flexible and interactive form of help by synthesizing knowledge, generating tailored responses, and offering task-specific guidance \cite{brynjolfssonGenerativeAIWork2025}. These capabilities may expand what workers can accomplish independently and thereby reduce their need for additional assistance \cite{dellacquaCyberneticTeammateField2026}. 
Moreover, privately seeking help from AI chatbots neither requires exposing one's own inability to another person nor does it incur reciprocity expectations \cite{nadlerEffectsOpportunityReciprocate1985}, potentially reducing the self-presentation costs that are key to the help-seeker's dilemma \cite{nadlerPersonalityHelpSeeking1997}. 
Consistent with this possibility, one recent study by Zhai and Qu \cite{zhaiCostAskingFit2027} found that seeking help from AI rather than human colleagues affords workers greater autonomy and reduces feelings of indebtedness, leading them to seek more help from AI in the future. Implications of using AI for work assistance may thus extend beyond the immediate helping exchange. 

Beyond being an alternative source of assistance, AI usage could also lead to \textit{downstream} effects --- impacting workers' subsequent help-seeking behaviour and sense of connection with their colleagues. This is especially relevant as AI is increasingly embedded \cite{brynjolfssonGenerativeAIWork2025}, and sometimes even organizationally mandated \cite{SeriousResultsUnserious,reporterTreasuryPushesShopifyInspired2025}, in how knowledge workers execute daily work tasks. Yu et al. \cite{yuUnveilingImpactEmployeeAI2026} surveyed service employees in China --- although not knowledge workers, they found that greater employee-AI collaboration predicted reduced help-seeking as a result of employees' increased psychological availability, or their perceived sufficiency of resources to carry out their work \cite{kahnPsychologicalConditionsPersonal1990}. Emerging evidence from educational settings also suggests that access to generative AI can reduce students' subsequent help-seeking from peers or instructors \cite{houAllRoadsLead2025,houEffectsGenerativeAI2024}. Although these studies were conducted outside knowledge work settings, their findings suggest that similar mechanisms could shape help-seeking in such settings.

Not only are the causal effects of working with AI on help-seeking unclear, but existing research has also largely treated AI use as a binary condition --- \textit{whether} it is being used or not --- but not \textit{how} AI is being used, despite known substantial variation in workers' AI usage \cite{balaynFrameworkCharacterizeReporting2026, bick2026rapid} and widespread gaps in organizational guidelines around AI \cite{WorkdayGlobalSurvey, zhangKnowledgeWorkersPerspectives2025}. In our paper, we experimentally examine both whether working with AI shapes subsequent help-seeking when compared to completing work without AI (\textbf{RQ1}), and whether this effect depends on key characteristics of how AI is being used (\textbf{RQ2}). In doing so, we offer a more detailed account of when and how using AI for work impacts help-seeking, and, consequently, workers' colleague relationships.   

\subsection{AI and Social Connections}

Much of the research on AI and social connections has been conducted outside the workplace, particularly in the context of therapy and companion chatbots \cite{taUserExperiencesSocial2020, skjuveMyChatbotCompanion2021, baebrandtzaegWhenSocialBecomes2021,mengMediatedSocialSupport2023}. Although these chatbots can provide immediate, non-judgmental support and reduce loneliness in the short term \cite{mengMediatedSocialSupport2023,defreitasAICompanionsReduce2026}, their use has also been associated with greater dependence and reduced human interaction \cite{laestadiusTooHumanNot2024,xieFriendMentorLover2023}, especially exacerbated among individuals with social anxiety or smaller social networks \cite{huHowSocialAnxiety2023,liuHeterogeneousEffectsAI2025}. Sycophantic AI may further undermine human connection by reducing willingness to repair interpersonal conflicts \cite{chengSycophanticAIDecreases2026,ibrahimSycophanticAIMakes2026}, and may encourage users to depend on frictionless AI interactions instead of more socially demanding, but meaningful human relationships. Altogether, prior work suggests that AI may offer immediate social support while simultaneously weakening long-term human connection.

Emerging workplace research around AI's broader social implications appears to be similarly mixed: frequent work-related AI use has been associated with employee loneliness and poorer well-being, including insomnia and increased alcohol consumption \cite{tangNoPersonIsland2023}. Yet, the same research found that AI use heightened employees' need for social affiliation, with prompted greater helping towards colleagues \cite{tangNoPersonIsland2023}. Moreover, loneliness may encourage help-seeking in workplace cultures characterized by high meaningfulness \cite{leeWhenLonelinessLeads2025}. Taken together, AI use may heighten workers' awareness of unmet social needs and motivate compensatory behaviours, including seeking or providing help. Another growing set of research focuses on how employees who use AI are being \textit{perceived} by others. Academic studies and industry reports suggest that workers who use AI --- or disclose doing so --- may be perceived as lazier \cite{doneganNewResearchShows2026,reifEvidenceSocialEvaluation}, less competent \cite{nakanoUnderstandingReaderPerception2026}, less trustworthy \cite{schilkeTransparencyDilemmaHow2025}, and less desirable to work with \cite{raeEffectsPerceivedAI2024,weissEffectsUsingArtificial2022}. For example, \textit{Atlassian} found that employees were less willing to recommend a colleague for a project when that colleague disclosed using AI even when the quality of their work is held constant, while also finding that organizational culture surrounding AI could mediate this social penalty \cite{doneganNewResearchShows2026}. Thus, it is fairly established that workers generally exhibit negative perceptions of colleagues who use or disclose using AI. 

In our paper, we flip the research emphasis from how colleagues \textit{perceive} AI-enabled workers to how these workers \textit{behave} toward their colleagues. Although perceptions can influence behaviour, limited research examines the social behaviours that follow AI use in the workplace. Studying help-seeking behaviour provides us a more direct account of potential social consequences as colleague relationships are shaped by workers' decisions to interact, coordinate, and engage --- or not --- with one another.

\section{Methodology}

We adopted the experimental vignette methodology \cite{aguinisBestPracticeRecommendations2014} to enable us to causally establish how a worker's AI usage for work tasks might shape their subsequent intended help-seeking behaviour, and, consequently, their sense of connection, towards their colleagues. Vignette studies have a long history of utilization in management literature \cite{eastmanEyesBeholderAttributional1994,bodeNoGoodGender2022,greenhausWhenWorkFamily2003}, including for help-seeking behaviours specifically \cite{liuAskingHowFish2022,vangerwenEmployeeCooperativeBehavior2018}, and has also been successfully executed in workplace HCI \cite{dasswainSensibleSensitiveAI2024, sadeghianSoulWorkEvaluation2024, chanInformingGroupInformatics2026}. This methodology presents participants with carefully crafted, realistic, hypothetical scenarios (i.e., vignettes) that encapsulate various factors of interest, for participants to assess based on their behaviours, intentions, and attitudes. The work context is especially suited to such an approach given the complexities of contemporary workplaces, which can make controlled field experiments infeasible and prohibitively expensive.
Vignettes also enable researchers to examine plausible near-future technologies, including various conditions of AI use, before they are widely deployed, by embedding hypothetical systems in recognizable and realistic work scenarios \cite{chanInformingGroupInformatics2026, lutzPrivacyConcernsSocial2021}.


\subsection{Hypotheses}

We developed five \textit{confirmatory} hypotheses (\textbf{H1---H5}) that map to each of our key factors of interest around AI usage, and how they impact colleague relationships. We start with \aipresence{} (\textbf{H1}), which distinguishes between \aienabled{} and \nonai{} vignettes to help us investigate the relational effects of using AI on work tasks. We then motivate each of the remaining factors and factor levels: \tasktype{} (\textbf{H2}), \role{} (\textbf{H3}), \specialization{} (\textbf{H4}), and \visibility{} (\textbf{H5}).

To operationalize colleague relationships for our hypotheses, we formulated two outcome variables, \dvhs{} and \dvcc{}, which were inspired and informed by the foundational Team-Member Exchange (TMX) construct \cite{seersTeammemberExchangeQuality1989, seersTeamMemberExchangeTeam1995, chenLiteratureReviewTeamMember2018}, designed to measure the quality of the working relationship between an individual and their other team members. Seers' \cite{seersTeammemberExchangeQuality1989} original TMX instrument included dimensions such as \textit{Exchange}, which assesses the extent to which team members give and receive feedback, ideas, knowledge, or assistance, as well as \textit{Cohesiveness}, which concerns members' sense of connection or belonging to the team. We were loosely inspired by the dimensions of \textit{Exchange} and \textit{Cohesiveness} to develop \dvhs{} and \dvcc{}, respectively. \dvhs{} captures workers' help-seeking intentions, and was conceptually included to represent a more proximal outcome. \dvcc{} was included to capture a broader and more distal social outcome, concerning workers' comfort, openness, and willingness to collaborate with a target colleague. 


\subsubsection{Factor Selection}

We prioritized and selected factors based on if they (1) concerned AI usage decisions over which individual workers or organizations have meaningful agency over, therefore having implications for practice or policy, and (2) reflected relevant forms of AI usage in the current and emerging workplace, bolstered by relevant literature. These factors were not intended to form a comprehensive taxonomy of workplace AI use. Other potentially important aspects of AI usage fell outside this scope. For example, AI tone --- while relevant to research around sycophantic AI \cite{chengSycophanticAIDecreases2026} --- was deliberately excluded, as default conversational behaviour is largely shaped by model providers. While users do have some level of agency indirectly through settings or prompting, we consider the range of AI prompting strategies to be fine-grained interaction behaviours rather than a broader factor of workplace AI usage. 

\subsubsection[AI Presence]{\aipresence[AI Presence]} 

We first examine whether \aienabled{} work affects colleague relationships relative to \nonai{} work, providing a baseline for evaluating the consequences on colleague relationships when introducing AI. 

Prior research has often contrasted seeking help from AI with seeking help from other people \cite{zhaiCostAskingFit2027}. We instead examine whether using AI to complete a task at work affects workers' subsequent help-seeking intentions towards their colleagues. Workers who use AI may feel less need to seek human assistance as AI could feel sufficient to address the problem at hand. Being readily available, AI also entails fewer social and psychological costs than approaching a colleague. This could reduce interaction opportunities and, consequently, colleague connection. We therefore hypothesize:

\begin{itemize}
\item \textbf{H1a:} \aienabled{} work is associated with lower \dvhs{} than \nonai{} work.
\item \textbf{H1b:} \aienabled{} work is associated with lower \dvcc{} than \nonai{} work.
\end{itemize}

\subsubsection[Task Type]{\tasktype[Task Type]}

The value of colleague assistance may depend on the task being performed. We developed four task types --- \assurance[Assurance], \brainstorming[Brainstorming], \proofing[Proofing], and \search[Search] --- that represent common workplace needs and can be completed with and without AI support. McGrath's circumplex model of group tasks, which organizes tasks around \textit{generating}, \textit{choosing}, \textit{negotiating}, and \textit{executing} activities \cite{polleyGroupsInteractionPerformance1984, mcgrathTimeInteractionPerformance1991}, as well as research on social support that distinguishes informational, instrumental, emotional, and esteem support \cite{cutrona1990type, schaeferHealthrelatedFunctionsSocial1981}, informed our development of the following task types. 

\assurance[Assurance] involves seeking emotional or esteem support that validates the worker's perspective surrounding a task. \brainstorming[Brainstorming] is a creativity task involving the generation of ideas. \proofing[Proofing] involves seeking feedback on the work task itself rather than validation of the worker's perspective. Finally, \search[Search] involves acquiring information relevant to the task and is a form of informational social support. AI tools may be able to more sufficiently address tasks like \proofing{}, and in particular, \search{}, that are more objective and well-defined, potentially reducing the need for subsequent help from colleagues. On the other hand, \assurance{} tasks may be perceived by workers to uniquely require a human touch. We thus expect task types to produce different levels of subsequent help-seeking and colleague connection, leading to our hypotheses: 

\begin{itemize}
    \item \textbf{H2a:} \tasktype[Task Type] is associated with \dvhs{}.
    \item \textbf{H2b:} \tasktype[Task Type] is associated with \dvcc{}.
\end{itemize}

\subsubsection[AI Role]{\role[AI Role]}

Help-seeking research distinguishes between instrumental help, where assistance focuses on offering ways to help solve problems, and executive help, which provides the solution itself \cite{karabenickRelationshipAcademicHelp1991,gallHelpseekingUnderstudiedProblemsolving1981}. Organizational research also draws a similar distinction between autonomous and dependent forms of help-seeking \cite{nadler1997autonomous}. We apply this distinction to the role workers have AI play in task completion. When AI is used in an \assisting{} manner (akin to instrumental or autonomous help), workers use AI in a supportive manner where they ultimately apply their own judgment and reasoning to the task. When AI is used in a \delegating{} manner (akin to executive or dependent help), workers use AI to complete the task, offloading responsibility to the AI. Although prior research has examined how these different approaches affect productivity or learning outcomes \cite{fongAcademicHelpseekingAchievement2023,gellerImpactHelpSeeking2012}, we examine their consequences on colleague relationships. 

For example, using AI in a \delegating{} manner may make workers' feel like further human input is less necessary as AI has already produced a solution that looks polished and complete, reducing subsequent help-seeking. In addition, the social risks of delegation may be relevant to aforementioned concerns about ``workslop'' \cite{liebscherWorkslopExaminingPrevalence}, as evidence has shown that this kind of output can harm recipient's perceptions of the sender. We therefore hypothesize: 

\begin{itemize}
    \item \textbf{H3a:} \role[AI Role] is associated with \dvhs{}.
    \item \textbf{H3b:} \role[AI Role] is associated with \dvcc{}.
\end{itemize}

\subsubsection[AI Specialization]{\specialization[AI Specialization]}

AI tools can vary in the knowledge available to them. Here, we compare \openassist{}, which draws primarily on generic knowledge and is meant to represent an off-the-shelf, but state-of-the-art AI chatbot like ChatGPT --- to \corpassist{}, which builds on \openassist{} but also incorporates organization-specific information (e.g., internal documents, past projects, team files, team communication) and tailors its responses accordingly. Organizations are increasingly developing specialized AI tools such as \corpassist{} to improve output relevance, leverage proprietary unstructured data, and protect sensitive information. A more specialized AI system may shape social behaviour as \corpassist{} could be perceived as being able to provide information that workers might have otherwise only been able to obtain from colleagues. However, workers may also view \corpassist{}, despite having access to organization-specific information, as insufficiently capable of capturing the tacit and contextual knowledge that is held by human colleagues. Examining AI specialization therefore reveals whether access to formal organizational knowledge is sufficient to displace colleague interactions. We hypothesize: 

\begin{itemize}
    \item \textbf{H4a:} \specialization[AI Specialization] is associated with \dvhs{}.
    \item \textbf{H4b:} \specialization[AI Specialization] is associated with \dvcc{}.
\end{itemize}

\subsubsection[AI Visibility]{\visibility[AI Visibility]}

A key element of the help-seeker's dilemma is the threat to self-esteem and self-presentation concerns. Because interactions with AI are typically private, workers can seek assistance without exposing their own uncertainties, incurring indebtedness, or risking judgment from colleagues \cite{nadlerPersonalityHelpSeeking1997,zhaiCostAskingFit2027}. However, AI interactions are increasingly being embedded into shared workplace platforms (e.g., Claude Tag\footnote{https://www.anthropic.com/news/introducing-claude-tag}) where conversations are both visible and accessible to colleagues. Visibility could change both the social and practical costs of subsequent help-seeking. By giving colleagues access to the worker's question, context, and history of interaction with the AI system, it could facilitate further discussion and shared understanding. On the other hand, workers may view a direct request as unnecessary because their need for assistance is already visible and colleagues can intervene if needed. Public AI use could also heighten self-presentation concerns, although visibility can sometimes provide an opportunity to demonstrate expertise through the questions asked. Given these competing possibilities, we hypothesize:

\begin{itemize}
    \item \textbf{H5a:} \visibility[AI Visibility] is associated with \dvhs{}.
    \item \textbf{H5b:} \visibility[AI Visibility] is associated with \dvcc{}.
\end{itemize}

\begin{figure}[h]
    \centering
    \includegraphics[width=\linewidth]{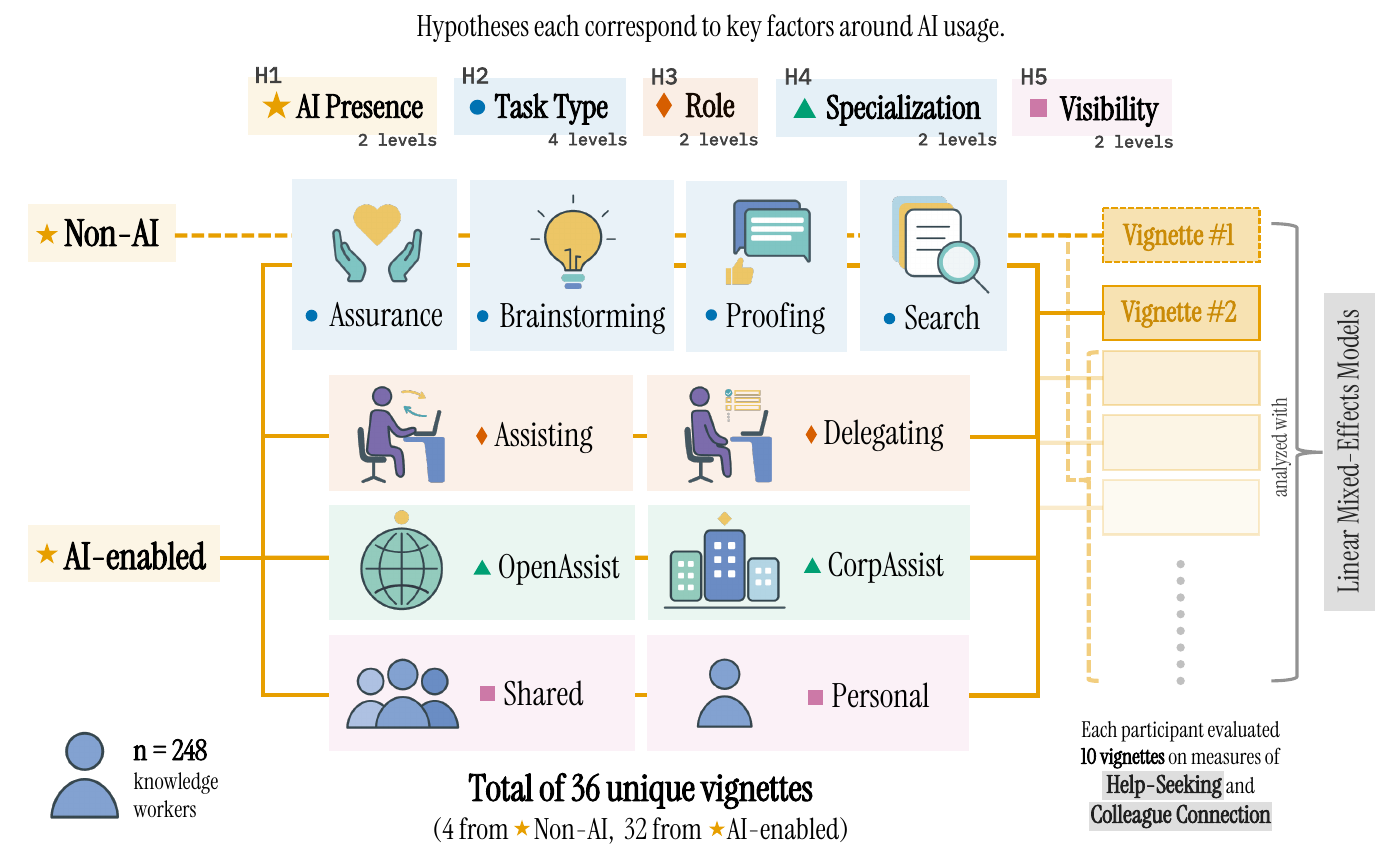}
    \caption{Overview of our experimental vignette study, including five key factors, each corresponding to one of the study's hypotheses. Overall, we had a total of 36 unique vignettes, of which participants evaluated a set of 10 vignettes each on \dvhs{} and \dvcc{}. We had one \nonai{} (dashed line) vignette per \tasktype{}, leading to 4 total. The 32 \aienabled{} (solid line) vignettes were constructed by combining \tasktype{} (4) $\times$ \role{} (2) $\times$ \specialization{} (2) $\times$ \visibility{} (2).}
    \Description{TBD}
    \label{fig:experiment}
\end{figure}

\subsection{Experimental Design}

As seen in Figure~\ref{fig:experiment}, we conducted a within-subjects factorial vignette experiment \cite{aguinisBestPracticeRecommendations2014} in which each participant evaluated a subset, or deck, of the complete set of 36 vignettes: 32 \aienabled{} conditions (4 levels of \tasktype{} $\times$ 2 of \role{} $\times$ 2 of  \specialization{} $\times$ 2 of \visibility{}) and 4 \nonai{} (1 per each level of \tasktype{}) conditions. To limit participant fatigue, each participant evaluated 10 vignettes: eight drawn from the \aienabled{} conditions and two from the \nonai{} conditions. Vignettes were assigned from a pre-computed counterbalanced schedule covering the 32-condition full factorial (eight AI vignettes per participant), so that each condition received approximately equal representation across the full sample. 
This design enabled within-participant comparisons of \aienabled{} and \nonai{} scenarios (\textbf{RQ1}) and of different configurations of AI usage (\textbf{RQ2}), while distributing coverage of the full factorial across our participant sample.

\subsubsection{Designing Vignettes}

We developed our vignettes iteratively to ensure that each clearly communicated the relevant factors, were grounded in an organizational context and realistic, and presented a situation in which either seeking or not seeking subsequent help were both plausible options. Each \aienabled{} vignette followed a consistent sequence: it first described \tasktype{}, followed by \specialization{} and \visibility{}, and concluded with \role{}. For \role{}, the description of how AI was used (either \assisting{} and \delegating{}) was tailored to each of the specific \tasktype{}. Each \nonai{} vignette similarly combined the description of \tasktype{} with text specific to each task explaining how ``you'' carried it out without AI assistance. Appendix Table~\ref{apx_tab:vignettes} presents the complete set of textual components used to construct the vignettes at each factor level. Each vignette also included visual iconography and highlighting to direct participants' attention to key information and help them recognize differences across vignette evaluations. The icons are reproduced in Appendix Figure~\ref{apx_fig:factor_icons}.

Before viewing the experimental vignettes, participants read a background vignette intended to establish and immerse participants into the broader organizational scenario. Here, we provided information about the organization, ``your'' role, the kinds of tasks ``you'' performed, ``your'' team and colleagues, and the organization's general climate of collaboration. To hold the nature of ``your'' relationship to ``your'' colleagues constant, we described the colleagues as peer workers (e.g., held similar context about the team's work and held similar tenure) with whom ``you'' have established professional working relationships. The background vignette also introduced the organization's AI tool, \texttt{Assist}, including its two versions: \openassist{}, a general-purpose version, and \corpassist{}, a company-specific version. It further explained the distinction between using \texttt{Assist} in a \personal{} versus a \shared{} manner. Visuals (see Appendix Figures~\ref{apx_fig:specialization_bg} and ~\ref{apx_fig:visibility_bg}) were designed and partially generated with ChatGPT to illustrate the \specialization{} and \visibility{} factors, particularly for participants who may not have personally encountered comparable AI tools or modes of use. 

\begin{figure}[h]
    \centering
    \includegraphics[width=\linewidth]{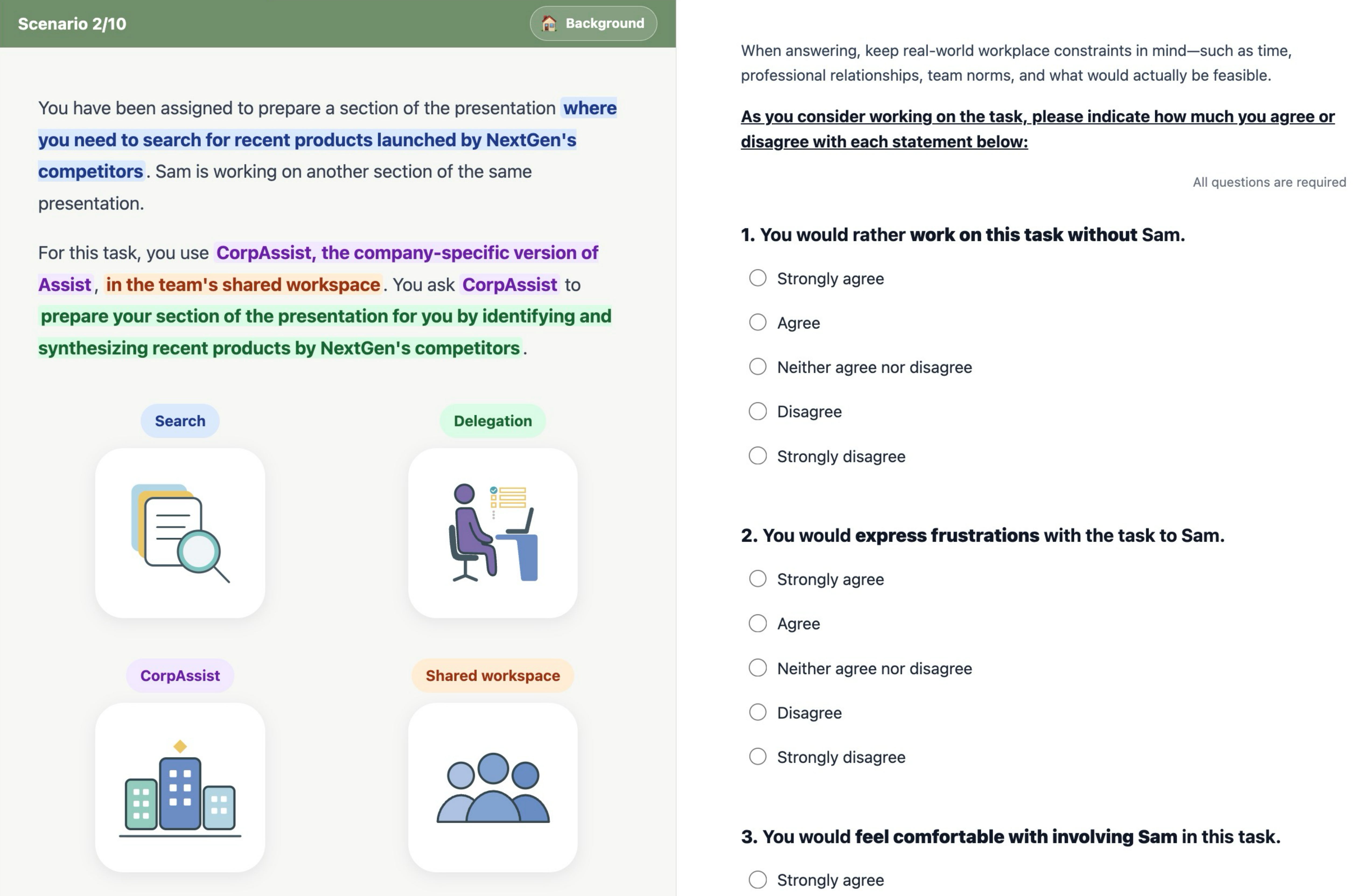}
    \caption{The online custom vignette portal where participants evaluated their set of ten vignettes. Visual iconography and highlighting is utilized to facilitate participant comprehension across vignettes. The panel on the right is scrollable to view all questions and proceed to the next vignette. Participants were required to access the portal only on a desktop or laptop computer.}
    \Description{TBD}
    \label{fig:portal}
\end{figure}

\subsubsection{Procedure}

After providing informed consent, participants completed a brief orientation on the online custom vignette portal (see Figure~\ref{fig:portal} for a screenshot of a vignette being evaluated) which presented the workplace background across multiple illustrated pages, with comprehension checks inserted between each page. When evaluating vignettes, participants were reminded to keep real-world workplace constraints in mind --- such as an imminent deadline (e.g., the tasks in the vignettes centred around a presentation due the same day), professional relationships, and what would be practically feasible. Attention checks were inserted at randomly selected positions among response items. 

Upon evaluating all ten vignettes in the portal, participants were redirected to complete the post-study survey in Qualtrics, which collected additional demographic information and administered various validated scales to include as covariates in our analyses, as we will discuss in Section~\ref{sec:data_analysis}. To contextualize why certain vignettes increased or decreased help-seeking intentions, participants also responded to two open-ended questions about the vignettes they just evaluated: \textit{``In the scenarios you just read, think of one scenario in which you were \{\textbf{MORE}/\textbf{LESS}\} likely to seek help from your colleague. Please explain your rationale in a few sentences.''}

\subsubsection{Expert Evaluation and Piloting}

To help refine our study design, we conducted hour-long Zoom interviews with six experts from industry and academia, including both researchers and practitioners with a range of backgrounds in the future of work, AI, developer experience, and organizational behaviour. Their feedback helped us concretize the AI usage factors examined in our study, identify relevant theoretical foundations and potential moderators, and iteratively refine our vignettes to ensure that each manipulation mapped clearly onto hypothesized changes in our outcome variables.
In particular, after experts highlighted esteem support as a key aspect of social support at work, we added \assurance{} as a new level of \tasktype{}. After our expert evaluations, we then also conducted three waves of pilot testing ($n=24$ total) to ensure the robustness of data collection through our study portal.

\subsection{Participants}

We recruited participants via Prolific, filtering for English-fluent, full-time workers from the United States with a 95\%+ study approval rate. Participants were paid \$10 USD for completing the study, which took a median of 20 minutes. The study was approved by the authors' ethics review board. Participants first completed an eligibility survey on Qualtrics that would then redirect them to our online vignette portal if found to be eligible. In addition to Prolific's prescreening filters, eligible participants needed to be knowledge workers who (1) used AI tools at work a few times a week or more, (2) did not have a lack of trust in AI tools' reliability, safety, and accuracy, (3) did not have supervisory responsibilities for more than one individual, and (4) collaborated with at least two or more colleagues on a typical work day. In other words, we wanted knowledge workers who could resonate most strongly with the workers in our vignettes --- who often collaborated with their colleagues, was comfortable with using AI for their work, and were mostly engaged in executing tasks rather than supervisory duties. 

Before finalizing recruitment, we conducted a power analysis for our planned linear mixed-effects models using the \texttt{simr} package in \textit{R} \cite{green2016simr}. We specified a significance level of $\alpha=0.05$ and a target effect size of $d=0.25$, which we converted to a fixed-effect coefficient ($b=0.182$) by multiplying the residual standard deviation ($\sigma_{pilot}=0.729$) estimated from pilot data with the effect size. For a more conservative estimate, we based our simulation on the analyses of \tasktype{} which had four factor levels compared to others with two. We ran $n=1000$ simulations, finding that 200 participants (or 1600 \aienabled{} vignette evaluations, eight per participant) would provide at least 90\% power to detect the target effect.



In total, 248 knowledge worker participants completed our study (see Appendix Table~\ref{apx_tab:demographics} for full demographics overview), resulting in 2444 vignette evaluations (1969 \aienabled{} and 475 non-AI vignette evaluations\footnote{$n=20$ participants were presented with only one \nonai{} vignette in an initial study data collection wave, which we later expanded to two.}). These totals reflect the exclusion of 16 vignette evaluations (15 \aienabled{}, 1 \nonai{})  completed in less than 10 seconds, a threshold corresponding approximately to the first percentile of evaluation times and deemed unlikely to reflect sufficient engagement with the vignette. The median evaluation time was 37 seconds, with 90\% of the evaluations completed within approximately 89 seconds. Most participants worked as analysts/researchers (44\%), with others in software development (17\%), project management (15\%), or writing/content creation (11\%). Participants generally worked in larger, enterprise-sized organizations, and more than 70\% of our participants reported working in organizations that provided AI training, as well as encouraged AI use.



\subsection{Measuring \dvhs{} and \dvcc{}}

To generate questions for measuring \dvhs{} and \dvcc{}, we first looked at TMX's \cite{seersTeammemberExchangeQuality1989} \textit{Exchange} and \textit{Cohesiveness} dimensions for inspiration.  Because the established TMX scale measures ongoing relationships within existing teams, they could not be applied directly to our vignette study --- questions needed to be connected to the work task described in each vignette.
As an example, the \dvhs{} items were inspired by TMX questions concerning the exchange of feedback and assistance, such as \textit{``In busy situations, how often do other team members ask you to help out?''} and \textit{``Do other members of your team usually let you know when you do something that makes their jobs easier (or harder)?''} We translated these concepts into three questions around participants' intended behaviour around seeking help from their target helper colleague --- in this case, Sam\footnote{Across the vignettes, Sam is one of four possible target helper colleagues. Sam's name is retained here for readability, but the name shown to participants in the question corresponded to the target colleague in the vignette.}: 

\begin{itemize}
    \item \textbf{Q1:} You would value Sam's feedback on this task.
    \item \textbf{Q2:} You would seek Sam's feedback on this task.
    \item \textbf{Q3:} You would actively incorporate Sam's feedback on this task.
\end{itemize}

Together, \textbf{Q1---3} capture aspects of participants' anticipated engagement in the help-seeking process --- valuing, seeking, and finally incorporating feedback. Similar to \dvhs{}, \textbf{Q4---6} were inspired by elements of trust and connection in the \textit{Cohesiveness} dimension of the original TMX instrument, as well as team psychological safety \cite{edmondsonPsychologicalSafetyLearning1999}. They also needed to be adapted to be centred around the task in our vignette study: 

\begin{itemize}
    \item \textbf{Q4:} You would feel comfortable with involving Sam in this task.
    \item \textbf{Q5:} You would express frustrations with the task to Sam.  
    \item \textbf{Q6:} You would rather work on this task without Sam. \textit{(reverse-coded)}
\end{itemize}

Before constructing our models, we assessed the internal consistency of both question sets across the full dataset using Cronbach's alpha (\dvhs{}, \textbf{Q1–3}: $\alpha=0.86$; \dvcc{}, \textbf{Q4–6}: $\alpha=0.63$). The \dvcc{} items showed lower internal consistency than expected. In an alpha-if-item-deleted analysis, removing \textbf{Q5} increased alpha from $0.63$ to $0.75$, indicating that responses to \textbf{Q5} were less consistent with responses to the other \dvcc{} items. We excluded \textbf{Q5} from subsequent analyses, in line with established practice in scale refinement \cite{boatengBestPracticesDeveloping2018,xiaoRevisitingUsageAlpha2024}. This decision was also informed by the item's conceptual distinction from the remaining questions: \textbf{Q5}'s focus on expressing frustration may have introduced greater emotional sensitivity or additional ambiguity. With only two items remaining, we instead calculated the correlation between \textbf{Q4} and \textbf{Q6} ($r = 0.61$; Spearman-Brown coefficient $\rho = 0.76$).
We constructed continuous measures of \dvhs{} and \dvcc{} for our analyses models by averaging the Likert-scale responses within each question set, with \textbf{Q5} excluded. Both measures ranged from $1$ to $5$, with higher values indicating greater \dvhs{} and \dvcc{}. 

\subsection{Data Analysis}
\label{sec:data_analysis}

We organized our analysis around our two key research questions and primarily built four linear mixed-effects models ($M$) in total: two for each of \textbf{RQ1} ($M_{RQ1,HS}$ and $M_{RQ1,CC}$) and \textbf{RQ2} ($M_{RQ2,HS}$ and $M_{RQ2,CC}$). $HS$ and $CC$ subscripts represent \dvhs{} and \dvcc{}, our outcome variables ($Y$), respectively. Each vignette evaluation acted as a data point in our study, with the entire dataset of \aienabled{} and \nonai{} vignette evaluations used to estimate $M_{RQ1}$ models, and only the \aienabled{} evaluations for the $M_{RQ2}$ models. We fitted linear mixed-effects models in R using the \texttt{lmer} function from the \texttt{lmerTest} package, which extended the \texttt{lme4} modelling framework \cite{de2011estimation}. We selected theoretically meaningful reference levels, using the condition we expected to elicit greater \dvhs{} as the baseline for each factor: \tasktype{} --- \assurance{}; \role{} --- \assisting{}; \specialization{} --- \openassist{}; and \visibility{} --- \shared{}. We used \aienabled{} as the reference for \aipresence{} so that the main effects of the experimental factors were estimated within the \aienabled{} condition, which was our primary condition of interest.
We report the basic model equation for each \textbf{RQ} below (see Equations~\ref{eqn:rq1_model} for \textbf{RQ1} and ~\ref{eqn:rq2_model} for \textbf{RQ2}). These basic models also include the order of presentation of the vignettes ($order\_vig$) as an additional covariate and the participant as a random effect ($1|participant$). 

\begin{equation}
\label{eqn:rq1_model}
\begin{split} 
    Y_{d} \sim ai\_presence + task\_type + ai\_presence:task\_type + order\_vig + (1 | participant)\\
    d \in \{\mathrm{\dvhs{}}, \mathrm{\dvcc{}}\}
\end{split}
\end{equation}

\begin{equation}
\label{eqn:rq2_model}
\begin{split} 
    Y_{d} \sim task\_type + role + specialization + visibility + order\_vig + (1 | participant)\\
    d \in \{\mathrm{\dvhs{}}, \mathrm{\dvcc{}}\}
\end{split}
\end{equation}

We also included covariates to account for potential additional factors that might explain the variance in the model. These covariates were added incrementally to the basic model based on three sets, loosely organized as: (1) \textit{individual worker demographics} --- age, gender, job function, job role, tenure, number of direct reports, work arrangement, collaboration frequency, number of colleagues they regularly interact with; (2) \textit{organizational context} --- organization size, extent to which organization provides AI training, extent to which organization encourages AI use, team psychological safety (TPS-7 \cite{edmondsonPsychologicalSafetyLearning1999}); and (3) \textit{individual personality} --- social anxiety (Mini-SPIN \cite{connorMiniSPINBriefScreening2001}), Big-Five personality traits (BFS-10 \cite{rammstedtMeasuringPersonalityOne2007}), and feedback orientation, or how an individual responds to feedback (Feedback Orientation Scale \cite{linderbaumDevelopmentValidationFeedback2010}). The full models, which were all found to have the highest conditional $R^2$ (see Appendix Section~\ref{apx_sec:model_tables} for all tables), comprises all three sets of covariates, with corresponding equations that can be found in Appendix~\ref{apx_sec:models}.

The order of presentation of vignettes was randomized, except for the first and second vignettes. We always first showed a \nonai{} vignette (they could also serve as a warm-up as these vignettes are simpler and shorter), and then an \aienabled{} vignette second. The other \nonai{} vignette slotted randomly in the remaining positions. 
With vignette order as a covariate in our full model (for \textbf{RQ1}, with both \aienabled{} and \nonai{} vignettes), we saw a minor, non-significant trend with \dvhs{} ($b=-0.009,p=0.068$), but not \dvcc{} ($b=-0.007,p=0.208$). This suggests that with each successive vignette, participants reported very slightly lower \dvhs{} intentions. We speculate that this may be due to fatigue, which accumulated as participants evaluated all ten vignettes, but the estimated difference per vignette was small.

\subsubsection{Analyzing Open-Ended Responses}

Where appropriate, we drew selectively on responses to our open-ended questions to supplement our quantitative findings. Two members of our research team independently and deductively coded each response, one using a dictionary-based classifier and the other using an LLM-assisted procedure (Anthropic's Claude Fable 5) in which the model was provided with descriptions of each of the AI usage factors, factor level definitions, and corresponding vignette wording. The LLM-assisted procedure was first calibrated against a manually coded subsample. The researchers then reviewed and resolved discrepancies between their coding results. We used the resulting classifications to identify candidate excerpts, which the researchers verified before including them as quotes. This analysis was intended to aid and enrich interpretation rather than serve as a standalone quantitative analysis, and thus should be interpreted as illustrations of potential explanations of the observed patterns. 

\section{Results}

Our results show that colleague relationships are indeed eroded when AI is involved in helping workers complete tasks (\textbf{RQ1}), and that its effects vary across different \tasktype[Task Types], with \brainstorming{} tasks being most negatively impacted. For \textbf{RQ2}, only \visibility{} had a significant association with the strength of colleague relationships. \majorr{We found limited evidence regarding the main effects of \role{} or \specialization{}.}


Results presented here are based on the full models, including our factors and all sets of covariates. The full models had the highest conditional $R^{2}$ and we observed no notable changes in the significance among our key factors. Here, conditional $R^2$ captures the variance explained by the full model, including both fixed effects and participant-level random effects. These were $M_{RQ1,HS}$ ($R^2=0.431$) and $M_{RQ1,CC}$ ($R^2=0.395$) for \textbf{RQ1}, containing the full set of vignette evaluations – both \nonai{} and \aienabled{}, as well as $M_{RQ2,HS}$ ($R^2=0.478$) and $M_{RQ2,CC}$ ($R^2=0.436$) for \textbf{RQ2}, which was fit only on the \aienabled{} vignette evaluations. 

We used a significance threshold of $\alpha=0.05$ to confirm or reject our hypotheses, while also noting trends ($p<0.10$) in our writeup as exploratory findings. We report unstandardized coefficient estimates, preserving the original Likert-scale units for ease of interpretation. For factors with more than two levels (e.g., \tasktype{}), we report results of omnibus $F$-tests followed by pairwise contrasts with Tukey-adjusted $p$-values; reported means are estimated marginal means. We report the fixed-effect coefficient directly for two-level factors. For brevity, we refer to our outcome variables \dvhs{} and \dvcc{} collectively as ``colleague relationship scores'', especially when both outcomes are broadly affected similarly. We round out our results by reporting on the effects of individual and organizational covariates.  

\subsection{Comparing Colleague Relationships across AI-Enabled and Non-AI Work Tasks}

\begin{figure}[t]
    \centering
    \begin{subfigure}[t]{0.48\linewidth}
        \centering
        \includegraphics[width=\linewidth]{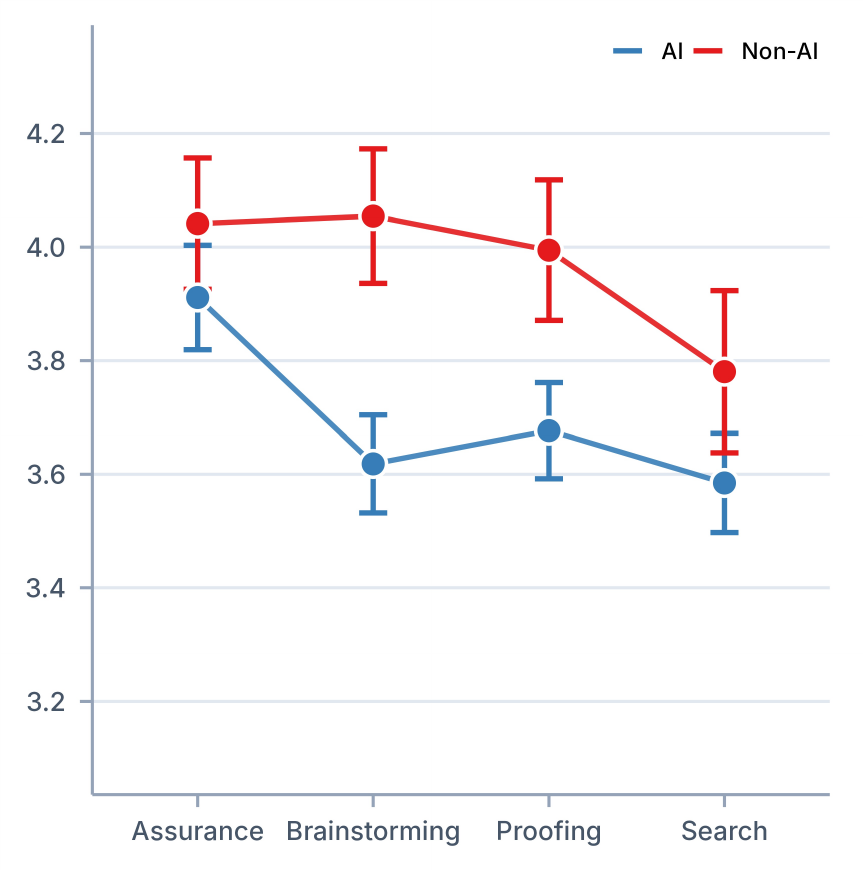}
        \caption{\dvhs{} means across different \tasktype[Task Types] and between \aienabled{} and \nonai{} vignettes.}
        \label{fig:rq1a}
    \end{subfigure}
    \hfill
    \begin{subfigure}[t]{0.48\linewidth}
        \centering
        \includegraphics[width=\linewidth]{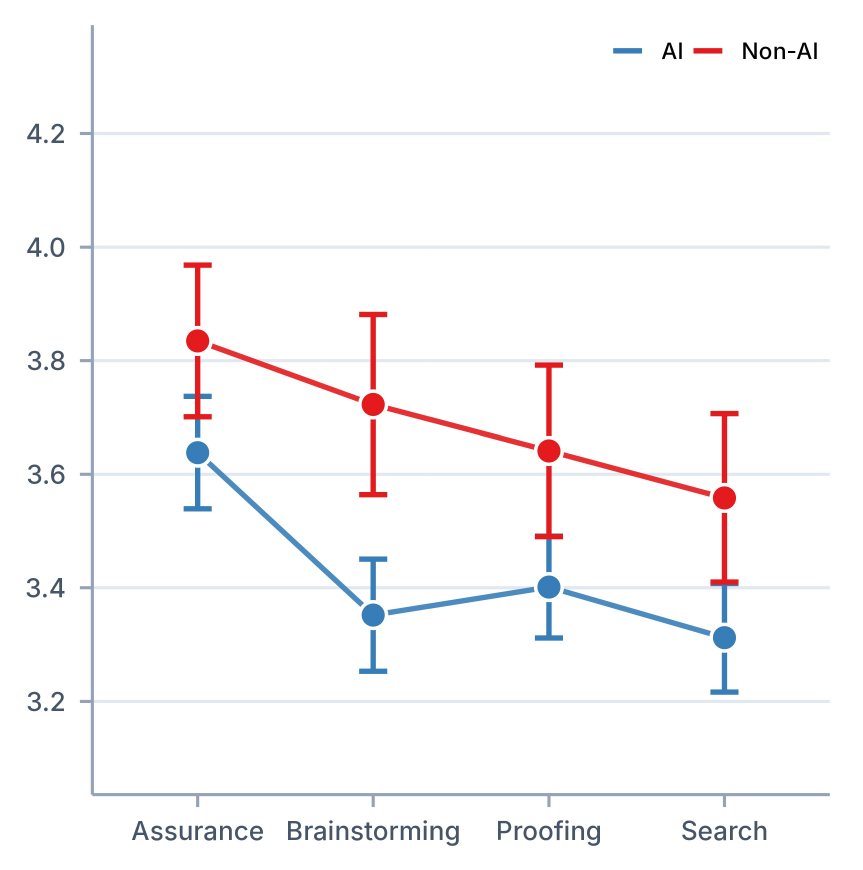}
        \caption{\dvcc{} means across different \tasktype[Task Types] and between \aienabled{} and \nonai{} vignettes.}
        \label{fig:rq1b}
    \end{subfigure}

    \caption{Both \dvhs{} and \dvcc{} are reduced in \aienabled{} vignettes, with the differences in only \dvhs{} being significantly lower in \brainstorming{} ($\Delta=-0.41$) and \proofing{} ($\Delta=-0.26$), with the \majorr{largest reduction in the former.}}
    \label{fig:rq1}
\end{figure}

We first asked the question of whether \aienabled{} work impacted colleague relationships across different task types (\textbf{RQ1}; see Table~\ref{tab:rq1}). As depicted in Figure~\ref{fig:rq1}, we looked at the main effects of \aipresence[AI Presence] and \tasktype[Task Type] along with any interaction effects between them for both colleague relationship scores. The full table of model results can be found in Appendix Tables~\ref{apx_tab:rq1_full_hs} and ~\ref{apx_tab:rq1_full_cc}.

\subsubsection{H1: AI-Enabled Work Leads to Weaker Colleague Relationships}
\label{sec:results_h1}

With \aienabled{} vignettes as a reference, we observed significantly higher \dvhs{} ($b=0.166,p=0.014$) and \dvcc{} ($b=0.233,p=0.003$) in \nonai{} vignettes, suggesting that AI use is associated with weaker colleague relationship outcomes. When AI is introduced to help with work tasks, across all vignettes, \dvhs{} is reduced by 0.240 points on average ($SE=0.036$), and \dvcc{} is reduced by 0.247 points on average ($SE=0.041$). Our findings lead us to reject the null hypothesis and that \textbf{H1 is supported} --- using AI for work tasks does indeed result in weaker colleague relationships, on the basis of reduced \dvhs{} and \dvcc{}. Open-ended responses suggest that weakened \dvhs{} intentions could be due to workers' reduced perceived need for help when using AI: \textit{``It's not that I wouldn't value the colleague assistance. I would, however, need it less --- I can work with the AI to get what I need and fact check its responses.''} Much of this pattern may hinge on \tasktype{}: \textit{``...less likely to [seek] help when the AI was able to complete most of the task for me. There was no additional help needed from my colleague. The AI provided enough support.''} In other words, if the nature of the task led participants to perceive AI as being capable of providing sufficient assistance, this could reduce their perceived need to consult colleagues, thereby reducing \dvhs{}.

\begin{table}[h]
\caption{\textbf{Summary of key results for RQ1}: estimated effects of \aipresence{}, \tasktype{}, and their interaction \aipresence{} $\times$ \tasktype{} from our linear mixed-effects models for \dvhs{} ($M_{RQ1,HS}$) and \dvcc{} ($M_{RQ1,CC}$). Table entries report coefficient estimates and corresponding $p$-values ($\cdot\ p < 0.10$, * $p < 0.05$, ** $p < 0.01$, *** $p < 0.001$). We observe significant main effects of \aipresence{} and \tasktype{}, as well as an interaction between \aipresence{} and \brainstorming{} in $M_{RQ1,HS}$.}
\label{tab:rq1}
\begin{tabular}{ll|lll|lll}
\multicolumn{2}{l|}{\multirow{2}{*}{\textbf{}}}
  & \multicolumn{3}{c|}{$M_{RQ1,HS}$}
  & \multicolumn{3}{c}{$M_{RQ1,CC}$} \\
\multicolumn{2}{l|}{}
  & \multicolumn{1}{c}{est.}
  & \multicolumn{2}{c|}{$p$-value}
  & \multicolumn{1}{c}{est.}
  & \multicolumn{2}{c}{$p$-value} \\ \hline

\multicolumn{1}{l|}{
  \multirow{2}{*}{
    \begin{tabular}[c]{@{}l@{}}
      \aipresence{}\\
      (ref: \aienabled{})
    \end{tabular}
  }
}
  & \nonai{}
  & 0.166 & 0.014 & \multicolumn{1}{c|}{*}
  & 0.233 & 0.003 & \multicolumn{1}{c}{**} \\
\multicolumn{1}{l|}{}
  & & & & & & & \\ \hline

\multicolumn{1}{l|}{
  \multirow{3}{*}{
    \begin{tabular}[c]{@{}l@{}}
      \tasktype[Task Type]\\
      (ref: \assurance{})
    \end{tabular}
  }
}
  & \brainstorming{}
  & -0.295 & \textless{}.001 & \multicolumn{1}{c|}{***}
  & -0.285 & \textless{}.001 & \multicolumn{1}{c}{***} \\
\multicolumn{1}{l|}{}
  & \proofing{}
  & -0.234 & \textless{}.001 & \multicolumn{1}{c|}{***}
  & -0.237 & \textless{}.001 & \multicolumn{1}{c}{***} \\
\multicolumn{1}{l|}{}
  & \search{}
  & -0.327 & \textless{}.001 & \multicolumn{1}{c|}{***}
  & -0.330 & \textless{}.001 & \multicolumn{1}{c}{***} \\ \hline

\multicolumn{1}{l|}{
  \multirow{3}{*}{
    \begin{tabular}[c]{@{}l@{}}
      \aipresence{} $\times$ \tasktype{}\\
      (ref: \aienabled{} $\times$ \assurance{})
    \end{tabular}
  }
}
  & \nonai{} $\times$ \brainstorming{}
  & 0.247 & 0.010 & \multicolumn{1}{c|}{**}
  & 0.155 & 0.159 & \multicolumn{1}{c}{} \\
\multicolumn{1}{l|}{}
  & \nonai{} $\times$ \proofing{}
  & 0.095 & 0.309 & \multicolumn{1}{c|}{}
  & -0.048 & 0.657 & \multicolumn{1}{c}{} \\
\multicolumn{1}{l|}{}
  & \nonai{} $\times$ \search{}
  & -0.044 & 0.638 & \multicolumn{1}{c|}{}
  & -0.050 & 0.646 & \multicolumn{1}{c}{} \\ \hline

\end{tabular}
\end{table}

\subsubsection{H2: Assurance Preserves Colleague Relationships, Whereas Brainstorming is Most Weakened}
\label{sec:results_h2}

We observed a significant main effect of \tasktype{} on \dvhs{} ($F=18.44, p<.001$) and \dvcc{} ($F=15.32, p<.001$), indicating that its impact on colleague relationships differed across \tasktype[Task Type] regardless of the presence of AI, making \textbf{H2 supported}. Overall, \assurance{} tasks had the highest mean colleague relationship scores (\dvhs{}: $3.97$; \dvcc{}: $3.78$), while \search{} tasks had the lowest (\dvhs{}: $3.62$; \dvcc{}: $3.42$). 
Relative to \assurance{}, all three other \tasktype[Task Types] were significantly lower in colleague relationship scores, suggesting that tasks where workers require their perspectives to be validated or affirmed by their colleagues were perceived to be ones that are most critical for \dvcc{} and that result in the most \dvhs{} --- intuitively, this means that \assurance{} tasks are overall more socially oriented. On the other hand, for \search{}, which had the lowest colleague relationship scores, participants highlighted its fact-based, objective nature, saying that as a result \textit{``discussing with my coworkers does not really change the outcomes much''}. They also mentioned the desire to not bother other colleagues with these kinds of requests, especially as AI can sufficiently help with them --- \textit{``...if I can get the answer myself... I'm a resourceful employee.''}


While colleague relationship scores were on average lower in \aienabled{} vignettes across all \tasktype[Task Types], we also found a significant \aipresence{} $\times$ \tasktype{} interaction only for \dvhs{} ($F=3.59, p=0.013$), indicating that the size of AI's impact on \dvhs{} varied somewhat across \tasktype[Task Types]. Pairwise comparisons of estimated marginal means showed that \aienabled{} vignettes were associated with significantly lower \dvhs{} for \brainstorming{} ($\Delta=-0.41, p<.001$)\footnote{$\Delta$ represents the difference between \aienabled{} and \nonai{} in \dvhs{}'s 5-point Likert-scale range.} and \proofing{} ($\Delta=-0.26, p=0.002$). The differences were smaller with \assurance{} ($\Delta=-0.17, p=0.219$) and \search{} ($\Delta=-0.12, p=0.620$) and were non-significant. Altogether, the results suggest that the relationship between \aipresence{} and \dvhs{} does vary by \tasktype[Task Type], with the largest reduction occurring for \brainstorming{} tasks. A possibility that we observed from the open-ended responses were that workers believed AI more capable of brainstorming than their colleagues: \textit{``I think I would yield better results from the AI than by asking my coworkers. I want organic ideas and I think trusting the AI is much more reliable and better sourced than my coworkers.''} This suggests that confidence in AI's output may outweigh and nullify the perceived benefits of human help-seeking --- even, interestingly, when generating ``organic'' ideas that might conventionally be associated with creative human input. 

Given its high colleague relationship scores and non-significant difference across \aipresence{}, participants also seemed to recognize the need for tacit knowledge associated with the \assurance{} task, and the limitations of AI in understanding that: \textit{``...the human input of a colleague who also experienced that project adds a nuanced context and some peer validation that a tool such as AI or a solo review cannot fully replicate.''} 


\subsection{How Different Ways of Using AI Impact Colleague Relationships}

After establishing that the presence of AI does indeed reduce colleague relationship scores and that these scores vary across tasks, we next describe how various AI usage factors shape colleague (\textbf{RQ2}; see Table~\ref{tab:rq2}). Here, like in \textbf{RQ1}, we similarly utilize linear mixed-effects models to test the main effects of AI-specific factors like \role[AI Role], \specialization[AI Specialization], and \visibility[AI Visibility], with full model results detailed in Appendix Tables~\ref{apx_tab:rq2_full_hs} and ~\ref{apx_tab:rq2_full_cc}.

\subsubsection{H3: No Significant Association Between AI Role and Colleague Relationships}
\label{sec:results_h3}

We found that \role[AI Role] (\assisting{} as reference compared to \delegating{}) did not significantly predict any changes in either \dvhs{} ($b=0.036,p=0.199$) or \dvcc{} ($b=0.034,p=0.290$), meaning that the null hypothesis sustains, and that consequently, \textbf{H3 is NOT supported}. We speculate that one plausible explanation is that both have AI \assisting[assist] you and \delegating[delegating] to AI sufficiently address workers' perceived need for help, even though they differ in how much of the task AI performs. Once that need is perceived as sufficiently met, how AI helps with the task may make little additional difference to whether workers perceive a need to seek help from a colleague.

\begin{table}[h]
\caption{\textbf{Summary of key results for RQ2}: estimated effects of \tasktype{}, \role{}, \specialization{}, and \visibility{} from our linear mixed-effects models for \dvhs{} ($M_{RQ2,HS}$) and \dvcc{} ($M_{RQ2,CC}$). Table entries report coefficient estimates and corresponding $p$-values ($\cdot\ p < 0.10$, * $p < 0.05$, ** $p < 0.01$, *** $p < 0.001$). We observe significant main effects of \tasktype{} and \visibility{} in both models.}
\label{tab:rq2}
\begin{tabular}{ll|lll|lll}
\multicolumn{2}{l|}{\multirow{2}{*}{\textbf{}}}
  & \multicolumn{3}{c|}{$M_{RQ2,HS}$}
  & \multicolumn{3}{c}{$M_{RQ2,CC}$} \\
\multicolumn{2}{l|}{}
  & \multicolumn{1}{c}{est.}
  & \multicolumn{2}{c|}{$p$-value}
  & \multicolumn{1}{c}{est.}
  & \multicolumn{2}{c}{$p$-value} \\ \hline

\multicolumn{1}{l|}{
  \multirow{3}{*}{
    \begin{tabular}[c]{@{}l@{}}
      \tasktype[Task Type]\\
      (ref: \assurance{})
    \end{tabular}
  }
}
  & \brainstorming{}
  & -0.295 & \textless{}.001 & \multicolumn{1}{c|}{***}
  & -0.286 & \textless{}.001 & \multicolumn{1}{c}{***} \\
\multicolumn{1}{l|}{}
  & \proofing{}
  & -0.234 & \textless{}.001 & \multicolumn{1}{c|}{***}
  & -0.237 & \textless{}.001 & \multicolumn{1}{c}{***} \\
\multicolumn{1}{l|}{}
  & \search{}
  & -0.328 & \textless{}.001 & \multicolumn{1}{c|}{***}
  & -0.332 & \textless{}.001 & \multicolumn{1}{c}{***} \\ \hline

\multicolumn{1}{l|}{
  \multirow{2}{*}{
    \begin{tabular}[c]{@{}l@{}}
      \role[Role]\\
      (ref: \assisting{})
    \end{tabular}
  }
}
  & \delegating{}
  & 0.036 & 0.199 & \multicolumn{1}{c|}{}
  & 0.034 & 0.290 & \multicolumn{1}{c}{} \\
\multicolumn{1}{l|}{}
  & & & & & & & \\ \hline

\multicolumn{1}{l|}{
  \multirow{2}{*}{
    \begin{tabular}[c]{@{}l@{}}
      \specialization[Specialization]\\
      (ref: \openassist{})
    \end{tabular}
  }
}
  & \corpassist{}
  & -0.026 & 0.353 & \multicolumn{1}{c|}{}
  & 0.006 & 0.860 & \multicolumn{1}{c}{} \\
\multicolumn{1}{l|}{}
  & & & & & & & \\ \hline

\multicolumn{1}{l|}{
  \multirow{2}{*}{
    \begin{tabular}[c]{@{}l@{}}
      \visibility[Visibility]\\
      (ref: \shared{})
    \end{tabular}
  }
}
  & \personal{}
  & -0.306 & \textless{}.001 & \multicolumn{1}{c|}{***}
  & -0.341 & \textless{}.001 & \multicolumn{1}{c}{***} \\
\multicolumn{1}{l|}{}
  & & & & & & & \\

\end{tabular}
\end{table}

\subsubsection{H4: No Main Effect of Specialization, But Exploratory Differences Across Task Types}
\label{sec:results_h4}

We found that \specialization[AI Specialization] (\openassist{} as reference compared to \corpassist{}) did not significantly predict any changes in either \dvhs{} ($b=-0.026,p=0.353$) or \dvcc{} ($b=0.006,p=0.860$), meaning that the null hypothesis sustains, and that consequently, \textbf{H4 is NOT supported}.

This finding was somewhat unexpected, as one could anticipate that a more specialized AI like \corpassist{} could impact workers' perceived need for subsequent help. We wondered if \tasktype{} could have played a role, as the additional context and knowledge from \corpassist{} maybe only mattered with particular tasks. We subsequently explored the interaction effect \tasktype{} $\times$ \specialization{}, finding a potential trend for \dvhs{} ($F=2.262,p=0.079$) but a significant interaction for \dvcc{} ($F=3.082, p=0.026$). Follow-up pairwise comparisons for \dvcc{} suggested that \corpassist{} may increase \dvcc{} ($b=0.156,p=0.022$) for \proofing{} but decrease it for \search{} ($b=-0.137,p=0.045$), relative to \openassist{}. The effect of \specialization{} also showed a trend toward differing between \proofing{} and \search{} ($b=0.293, p=0.062$).

The decrease in \search{} could be an intuitive result, as a more specialized AI may be perceived as being able to handle objective information-seeking work tasks better than a more generic one. However, it is less immediately clear why \corpassist{} is associated with greater \dvcc{} for \proofing{}.
One possible explanation is that participants perceived \corpassist{}'s access to organization-specific knowledge as insufficient in judging its relevant and appropriateness in the context of a \proofing{} task, which involves an understanding of when to actually \textit{apply} that knowledge. It may be that with this in mind, participants find colleagues more valuable for verifying \corpassist{}'s feedback.
Regardless, recognizing the exploratory nature and modest magnitude of these interaction effects, we interpret these patterns lightly as potential trends. 

\subsubsection{H5: Shared AI Workspaces Are Associated With Stronger Colleague Relationships}
\label{sec:results_h5}

\visibility{} had the strongest significant main effect out of the AI usage factors across both colleague relationship outcomes. Compared to using AI in \shared[shared], using AI in a \personal[personally] led to decreases in both \dvhs{} ($b=-0.306,p<.001$) and \dvcc{} ($b=-0.341,p<.001$), thus leading to \textbf{H5 being supported}. This suggests that when workers interact with AI in a workspace that is visible to their colleagues, they report greater help-seeking intentions towards their colleagues along with stronger connection to them than when the same AI interaction occurs privately. Notably, private AI interactions remain the predominant mode through which workers interact with AI tools in many organizations.

Participants described how the \shared[shared] workspace reduced both the practical and social costs of seeking help through several pathways. With a \shared[shared] workspace providing colleagues access to relevant context, participants felt like they could ask for help \textit{``without me having to explain it from scratch.''} It also made prior effort visible, making them more comfortable asking for help: \textit{``...I've proved to have tried already.''} Finally, the workspace reduced concerns around disclosing AI use because \textit{``...they already know that I'm using AI.''} By contrast, seeking help when using AI \personal[personally] \textit{``felt more vulnerable and like more of an imposition --- on their time and into my `space.'''} In this quote, we see concerns about burdening the helper, a well-established psychological cost of help-seeking, as well as a sense that using AI personally made the work process feel more private and costly to subsequently share with a colleague. Together, these quotes suggest plausible mechanisms through which a shared AI workspace may lower the psychological costs of seeking help. 

\subsection{Exploring Individual and Organizational Differences}
\label{sec:results_covariates}

We report \majorr{additional exploratory findings} for the various participant-level covariates comprising our full linear mixed-effects models. We focus on the models estimated using the \aienabled{} vignettes ($M_{RQ2,HS}$ and $M_{RQ2,CC}$; see Appendix Tables~\ref{apx_tab:rq2_full_hs} and ~\ref{apx_tab:rq2_full_cc}, respectively) to supplement our hypotheses and help further contextualize our results. 

\textbf{Organizational Setting and Work Arrangement Did Not Explain Colleague Relationship:} Most of the other covariates did not have statistically significant main effects on either of our colleague relationship outcomes. Among organizational factors, we found no significant differences associated with work arrangement (in-person, remote, or hybrid), organization size, or team psychological safety. We also found no significant differences based on whether participants' organizations provided training on or encouraged AI use.
These results \majorr{could hint} that differences in \dvhs{} or \dvcc{} were not primarily explained by the organizational factors included in our study, and that our observed effects of AI use on colleague relationships may span different work arrangements and organizations. However, this does not preclude the potential role of other organizational factors, practices, or even interventions in addressing AI's relational impact. 

\textbf{Feedback Orientation Has Weak Association with Help-Seeking, but Other Internal Traits Do Not:} Individual personality covariates in our study likewise did not significantly predict either \dvhs{} or \dvcc{}. In particular, social anxiety was not significantly associated with \dvhs{} ($p=0.174$) or \dvcc{} ($p=0.746$), nor was extraversion (\dvhs{}: $p=0.968$, \dvcc{}: $p=0.737$). We did see significantly increased \dvhs{} ($b=0.157,p=0.035$) for the \textit{Utility} subscale of the Feedback Orientation Scale \cite{linderbaumDevelopmentValidationFeedback2010}, which measures the extent to which an individual believes that receiving feedback leads to helpful outcomes. We observed no corresponding association in \dvcc{}.

\textbf{Individual Contributors' Colleague Relationships are More Vulnerable:}  We recruited knowledge workers with at most one direct report, i.e., did not primarily work in a supervisory role. Whether participants had zero or one direct reports did not have a significant effect on our outcomes (\dvhs{}: $p=0.149$; \dvcc{}: $p=0.504$). Having said that, 8.5\% ($n=21$) of our participants reported their job role as managers, recognizing that managerial roles can encompass responsibilities beyond direct people management. We compared them to knowledge workers who reported themselves to be junior individual contributors (ICs; labeled as ``skilled office worker'' --- 56\%, $n=138$) or senior ICs (36\%, $n=89$). Relative to managers, both junior and senior ICs reported lower \dvhs{} and \dvcc{}. Junior ICs scored $0.496$ points lower on \dvhs{} ($p=0.002$) and $0.398$ points lower on \dvcc{} ($p=0.019$). The differences were slightly less pronounced for senior ICs, who scored $0.411$ points lower on \dvhs{} ($p=0.009$) and $0.395$ points lower on \dvcc{} ($p=0.022$). These findings may suggest that workers in management or seniority roles, compared to more junior ICs, are more inclined to seek help and feel more connected to their colleagues, even when AI is being used. At the same time, we found no significant differences associated with tenure at participants' current organizations: compared with participants with less than two years of tenure, neither those with three to five years (\dvhs{}: $p=0.532$, \dvcc{}: $p=0.841$) nor those with six or more years (\dvhs{}: $p=0.974$, \dvcc{}: $p=0.457$) differed significantly on either outcome. Thus, we speculate that one possibility is that these differences reflect the nature of participants' work rather than seniority itself. Junior ICs may perform more execution-oriented tasks and fewer decision-making tasks, creating more opportunities for AI to reduce the perceived need to seek help from colleagues. This would align with how \tasktype{} was found to have a key impact on \dvhs{} and \dvcc{}.

\section{Discussion}

We show that AI use reduces workers' intentions to seek help from and feel connected to their colleagues, though the magnitude of this effect varies by task type and is greatest for brainstorming (\textbf{RQ1}). Shared AI workspaces may mitigate this decline, particularly relative to conventional, individual AI use (\textbf{RQ2}). 
Considered more broadly, our findings have implications for social well-being at work. Work is an important source of meaning and social connection, and relationships with colleagues can support worker well-being and buffer against stress \cite{colbertFlourishingWorkplaceRelationships2016,jollySocialSupportWork2021}. Yet, one in five workers experience loneliness at work \cite{incStateGlobalWorkplace2026}, and all the while AI is becoming increasingly embedded in workplaces. As our findings suggest that AI use may continue to erode colleague relationships, it is important to consider how the human relationships that support individual and organizational flourishing can be preserved. 

This may require moving beyond a narrow, efficiency-centric view of work, in which help-seeking may appear increasingly unnecessary as AI becomes more and more capable. Instead, more holistic perspectives of work \cite{guillouYourTimeWell2020} could help us reframe help-seeking in the age of AI not solely as a means of completing tasks or improving productivity, but as a \textit{relational practice} through which social capital \cite{nahapietSocialCapitalIntellectual1998} is fostered and sustained.
With appropriate practices, policies, norms, and continued attention to AI design, reduced help-seeking need not necessarily lead to detrimental social consequences if human help becomes more focused on situations where colleagues offer distinct value.
We adopt this broader perspective in the following sections as we interpret our findings and consider their implications. Specifically, we discuss how AI may shift the expected perceived benefits of human assistance, introduce new self-presentation costs, and complicate when and where human help is still needed. We conclude by reflecting on the value of shared AI workspaces, including their potential drawbacks.

\subsection{Shifting Expectations of the Benefits of Human Assistance}

One plausible explanation of our observed help-seeking erosion (\S~\ref{sec:results_h1}) was that AI reduced participants' perceived need for human help, thereby changing the relative benefit of turning to a colleague. 
Yet reduced need may be only one way that AI impacts relative benefit. We speculate on how other qualities of AI use at work might further shift the perceived benefits of seeking help from human colleagues.

``That's a \textit{great} question!'' Many of us have encountered similarly enthusiastic responses from AI, even when the question asked might have been ill-formed or simply mundane. AI chatbots are always capable of providing a polished, pleasant response, no matter how you prompt it for help. Over time, our expectations around receiving help from other human colleagues may plausibly shift as a result of our exposure to AI, and perhaps especially with sycophantic AI, which has already been shown to reduce prosocial intentions \cite{chengSycophanticAIDecreases2026}. We may come to expect the same kind of agreeableness from our colleagues as we have become so used to with AI. As AI --- the ultimate yes-man \cite{mukundaAIUltimateYesMan2025} --- will almost always attempt to cobble together an answer, we may even become dissatisfied when a human helper admits, ``I don't know'' --- viewing this as a \textit{shortcoming} of our colleague, rather than a possible \textit{feature}: a signal that the question requires additional resources, clarification, or deeper consideration. We may even come to expect the responsibility for understanding our help-seeking requests to lie with the target helper, rather than with us --- especially since AI will, of course, answer any question we ask it, no matter how many grammatical errors, typos, or ambiguities it contains. 
Future work could explore the plausibility of such a mechanism, focusing specifically on worker expectations of their colleagues, and how it might shape subsequent help-seeking. 
Finally, as AI output quality continues to improve, the relative benefit of seeking help from a human colleague may further diminish, potentially reducing help-seeking beyond what we observed in our study and making this an ongoing challenge with evolving AI capabilities.

\subsection{AI Adds New Self-Presentation Costs when Seeking Help}

The costs associated with seeking help traditionally were psychological and social, and revolved around \textit{self-presentation} --- our human desire to present ourselves favourably to others. In contemporary workplaces, where norms and attitudes around AI use still vary considerably across individuals and organizational contexts, AI may introduce new self-presentation costs to help-seeking, particularly around workers' desire to appear competent to their colleagues. Our open-ended responses (\S~\ref{sec:results_h5}) suggest that one possible explanation may be the discomfort associated with disclosing AI use when seeking help, which may compound the existing self-presentation costs of admitting the need for help in the first place \cite{bambergerEmployeeHelpseekingAntecedents2009}. 
Shame, fear, and avoidance around disclosing AI have been documented in prior work \cite{mizrakWhyHideAI2026, baoAIDisclosureMoral2026, hwangBetterAskForgiveness2026}, particularly as disclosure can lead to reduced trust and other negative social evaluations \cite{schilkeTransparencyDilemmaHow2025, reifEvidenceSocialEvaluation}.
Workers may therefore expend additional effort in masking AI-assisted work on a task before feeling comfortable with asking for help, especially as AI use raises questions about control and ownership over the work \cite{kadomaRoleInclusionControl2024}. In this way, AI may introduce an additional layer of self-presentation costs to the already socially-laden help-seeker's dilemma. 

Another plausible explanation is that workers desire to signal competence, independence, and effort (\S~\ref{sec:results_h5}) to their colleagues \cite{bambergerEmployeeHelpseekingAntecedents2009,brooksSmartPeopleAsk2015}, but AI may complicate how workers now approach signalling these qualities. In illuminating why shared AI workspaces had higher help-seeking, open-ended responses suggested that having the AI conversation be visible acts as a signal of the help-seeker's effort in attempting to resolve the problem on their own first. Yet, the utility of AI may change what an individual believes they \textit{should} be able to accomplish on their own. When web search became popular, \textit{Let Me Google That For You}\footnote{https://lmgtfy.com/}, emerged out of a frustration that people were not being resourceful but needlessly bothering others. AI drastically expands the range of what an individual could (or perceives they could) accomplish on their own\footnote{In fact, there is now a \textit{Let Me ChatGPT That For You}: https://lmcgpttfy.app/} --- and, as a result, may decrease human help-seeking out of a desire to be seen as competent. An opportunity for organizational leaders is to reinforce that ``AI could have answered that'' is not, in itself, a reason to avoid seeking help from colleagues, while remaining mindful of collaborative overload \cite{crossCollaborativeOverload2016} --- striking a balance between independence and interdependence. The challenge may then lie in understanding when a human touchpoint is required, which we only begin to touch on next. 


\subsection{The \textit{AI-Enabled Help-Seeker's Dilemma}: Discerning When Human Help Is Still Needed}

The preceding sections discuss mechanisms by which AI-enabled work reshapes both the perceived benefits and costs of seeking help from colleagues, central components of the classic \textit{help-seeker's dilemma} \cite{nadlerPersonalityHelpSeeking1997}. 
Revisiting and extending this dilemma in light of AI-embedded knowledge work, we posit the \textit{AI-enabled help-seeker's dilemma}, which offers a new challenge around discerning when additional human help is still needed, particularly when AI may seem to have already resolved the need for help. This is best illustrated by one of our key findings: AI decreased help-seeking most for brainstorming tasks (\S~\ref{sec:results_h2}). Brainstorming is distinct from more objective, information-seeking tasks because it can be difficult to judge whether an AI-generated response is sufficient \cite{limWorkslopWorkHow2026}. The polish of AI output may pose additional risks for novices, who may not yet have developed the ability to distinguish fluency from depth.
AI-generated ideas may appear complete while flattening information \cite{doshiGenerativeAIEnhances2024}, lacking the divergent and unexpected contributions that human colleagues can bring. Yet, this does not preclude a role for AI in augmenting human-human creativity, with prior HCI work demonstrating its potential to act as a ``social glue'' \cite{suhAISocialGlue2021}. This points to a design opportunity for AI tools that scaffold brainstorming without displacing human help-seeking, instead creating opportunities to bring colleagues into the creative process and combine both AI support with the divergent perspectives that colleagues can provide.

Our findings also suggest that tasks involving validation and perspective (categorized as \assurance{} in the study, \S~\ref{sec:results_h2}) may more clearly benefit from colleagues' tacit and organizational knowledge, which AI may not be able to fully capture. By contrast, for other tasks, sufficiently convincing AI output might obscure what is lost by not involving human others. Rather than seeking to preserve help-seeking uniformly across task types, there may be value in redistributing help-seeking, or perhaps framed more broadly, support interactions, towards situations where people offer distinctive value --- with the human-human connection considered meaningful especially in AI-embedded work \cite{ranjitAreWeAutomating2026}. This might mean a shift away from instrumental assistance towards emphasizing mentorship, perspective, and human connection. 


Altogether, a key implication for organizations is to move beyond broad AI training towards task-specific guidance about when human input remains valuable, and how to ask for help in an AI-embedded workplace. 
For example, teams could integrate human touchpoints within AI-enabled workflows, akin to code reviews in software development --- particularly for tasks like brainstorming, where assessing the sufficiency of AI output may be difficult and additional human perspectives valuable, and especially for novice workers.
Ultimately, what kinds of work still warrant human help in an AI-enabled workplace remains an open question in the AI-enabled help-seeker's dilemma --- here, we only begin to surface this tension in the context of help-seeking and encourage future work around this question.

\subsection{Shared AI Workspaces: A Double-Edged Sword?}

When considering how AI tools for work might be designed and used to preserve colleague relationships, one of our key findings is that shared, visible AI use may help sustain human help-seeking (\S~\ref{sec:results_h5}). The visibility of a shared AI workspace could make work easier to discuss with colleagues, creating opportunities for others to contribute or intervene and lowering the cost of help-seeking. It is worth acknowledging, however, that a shared AI workspace, as described in our vignettes, bypasses the psychological cost of admitting that you need help by making the problem and task already visible to colleagues. Future work could investigate workers' willingness to engage with shared AI workplaces in the first place, and how human-AI collaboration evolves in this emerging form of multi-user collaboration. 

Yet, the ease of sharing in visible AI workspaces may also come at a cost. After a lengthy back-and-forth with AI about a task, in such workspaces, a help-seeker can easily involve a colleague by tagging them into the conversation. However, this shifts much of the cost of seeking help onto the prospective helper, who must sift through the conversation, identify relevant context, before finally determining how to help. CSCW research has long emphasized the \textit{articulation work} involved in exchanging help \cite{crabtreePracticalIndispensabilityArticulation2006}. Shared AI may bypass this articulation work by making it particularly easy for help-seekers to offload that work solely onto colleague target helpers, which may simply result in another type of ``workslop'' \cite{liebscherWorkslopExaminingPrevalence}. Naively introducing shared AI workspaces without appropriate guardrails may only produce collaborative overload rather than meaningfully encouraging help-seeking and strengthening colleague work relationships. 
A possible design opportunity could be for shared AI workspaces to help summarize relevant context and help workers formulate clear, focused requests before involving a colleague --- in other words, \textit{having AI assist us in becoming better help-seekers}, rather than \textit{solely focusing on having it solve our need for help in the first place}. As an example, such a relational approach to AI design could build upon emerging HCI work demonstrating that AI optimized and designed to account for relationships between people can improve cooperation \cite{claggettRelationalAIFacilitating2025}.
Working openly with AI may reduce the stigma associated with speaking up and asking for help, especially with the aforementioned potential added costs when AI is being used. Regardless, we argue that the extent to which these benefits can be realized do ultimately rest on psychological safety and organizational culture --- not simply on the introduction of a new technological system. 

\subsection{Limitations and Future Work}

A common limitation of vignette studies is that they capture intended rather than actual behaviour, although much prior work has supported their value as approximations of real-world decision-making \cite{hainmuellerValidatingVignetteConjoint2015}. We also used generic knowledge work tasks in our vignettes to ensure evaluation validity across knowledge worker participants from various industries and occupations, potentially overlooking domain-specific norms and constraints. Future studies could examine actual help-seeking behaviour in simulated AI-enabled work environments, as well as target specialized groups of knowledge workers --- such as software developers --- whose distinct help-seeking needs may be particularly impacted by AI. While Prolific-based recruitment may also raise concerns about engagement, attention checks, human-verification and bot-detection procedures, as well as data-quality screening provide support for response quality.

We did not conduct mediation analyses because our research questions focused on whether --- and under what factors --- AI usage shapes help-seeking, rather than on the underlying mechanisms. Establishing such mechanisms may require a separate study design, potentially involving the manipulation of proposed mediators and sequentially separating the mediator and outcome measures for robustness. Future work could more directly test mechanisms that we speculate on based on our results, including the perceived social costs of disclosing AI use and the heightened expectations that workers should be able to solve problems independently when AI is available.

Our U.S.-only sample also limits cross-cultural generalizability. Although the individualism-collectivism distinction overly simplifies cultural variation \cite{fiskeUsingIndividualismCollectivism2002,oysermanRethinkingIndividualismCollectivism2002}, the patterns observed here may differ especially in more collectivist contexts, where workplace help-seeking, help-giving, and AI use may have distinct norms, expectations, and perhaps even obligations. Future work could target cross-cultural replication.

\section{Conclusion}

Strong colleague relationships are key to individuals' flourishing at work. The exchange of help is one way such relationships are fostered. 
Yet, our experimental vignette study showed that in AI-enabled work, workers' subsequent help-seeking intentions decreased. This social erosion varied across different work tasks, but making AI usage visible among colleagues may have helped retain help-seeking.
As organizations continue to grapple with how to integrate AI into the workplace while accounting for potential social consequences, our work empirically illuminates the extent to which, and the conditions under which, AI use may weaken or preserve colleague relationships, while surfacing opportunities for organizational action. In doing so, we optimistically envision an AI-enabled future of work that can foster more focused and meaningful help-seeking around the kinds of support and relationship that is uniquely human. 




\bibliographystyle{ACM-Reference-Format}
\bibliography{ref}

\appendix
\clearpage
\onecolumn
\section{Participant Demographics}
\begin{table}[htbp]
\centering
\caption{Overview of participant demographics ($n = 248$).}
\label{apx_tab:demographics}
\small
\renewcommand{\arraystretch}{0.9}
\begin{minipage}{0.86\textwidth}
\begin{minipage}[t]{0.48\linewidth}
\vspace{0pt}
\begin{tabular*}{\linewidth}{@{\extracolsep{\fill}}lr@{}}
\toprule
\textbf{Gender} & \textbf{n (\%)} \\
\midrule
Woman & 136 (54.8\%) \\
Man & 108 (43.5\%) \\
Other (Non-Binary) & 3 (1.2\%) \\
Other (Prefer Not to Disclose) & 1 (0.4\%) \\
\phantom{Placeholder} & \phantom{0 (0.0\%)} \\
\bottomrule
\end{tabular*}
\end{minipage}%
\hfill
\begin{minipage}[t]{0.48\linewidth}
\vspace{0pt}
\begin{tabular*}{\linewidth}{@{\extracolsep{\fill}}lr@{}}
\toprule
\textbf{Age} & \textbf{n (\%)} \\
\midrule
18–24 & 25 (10.1\%) \\
25–34 & 100 (40.3\%) \\
35–44 & 63 (25.4\%) \\
45–54 & 47 (19.0\%) \\
55–64 & 13 (5.2\%) \\
\bottomrule
\end{tabular*}
\end{minipage}
\smallskip
\begin{minipage}[t]{0.48\linewidth}
\vspace{0pt}
\begin{tabular*}{\linewidth}{@{\extracolsep{\fill}}lr@{}}
\toprule
\textbf{Tenure} & \textbf{n (\%)} \\
\midrule
$\leq$2 years & 85 (34.3\%) \\
3–5 years & 83 (33.5\%) \\
6+ years & 80 (32.3\%) \\
\phantom{Placeholder} & \phantom{0 (0.0\%)} \\
\phantom{Placeholder} & \phantom{0 (0.0\%)} \\
\bottomrule
\end{tabular*}
\end{minipage}%
\hfill
\begin{minipage}[t]{0.48\linewidth}
\vspace{0pt}
\begin{tabular*}{\linewidth}{@{\extracolsep{\fill}}lr@{}}
\toprule
\textbf{Organization Size} & \textbf{n (\%)} \\
\midrule
\textless{}50 & 25 (10.1\%) \\
50–249 & 41 (16.5\%) \\
250–999 & 41 (16.5\%) \\
1,000–9,999 & 76 (30.6\%) \\
10,000+ & 65 (26.2\%) \\
\bottomrule
\end{tabular*}
\end{minipage}
\smallskip
\begin{minipage}[t]{0.48\linewidth}
\vspace{0pt}
\begin{tabular*}{\linewidth}{@{\extracolsep{\fill}}lr@{}}
\toprule
\textbf{Work Arrangement} & \textbf{n (\%)} \\
\midrule
In-Person & 57 (23.0\%) \\
Hybrid & 104 (41.9\%) \\
Remote & 87 (35.1\%) \\
\bottomrule
\end{tabular*}
\end{minipage}%
\hfill
\begin{minipage}[t]{0.48\linewidth}
\vspace{0pt}
\begin{tabular*}{\linewidth}{@{\extracolsep{\fill}}lr@{}}
\toprule
\textbf{Job Role} & \textbf{n (\%)} \\
\midrule
Skilled Office Worker & 138 (55.6\%) \\
Senior Staff & 89 (35.9\%) \\
Managers & 21 (8.5\%) \\
\bottomrule
\end{tabular*}
\end{minipage}
\smallskip
\begin{minipage}[t]{\linewidth}
\vspace{0pt}
\begin{tabular*}{\linewidth}{@{\extracolsep{\fill}}lr@{}}
\toprule
\textbf{Job Function} & \textbf{n (\%)} \\
\midrule
Analysis / Research & 109 (44.0\%) \\
Software / Engineering & 42 (16.9\%) \\
Planning / Project Management & 37 (14.9\%) \\
Writing / Content Creation & 26 (10.5\%) \\
Management / Strategy & 18 (7.3\%) \\
Design / Creative Work & 16 (6.5\%) \\
\bottomrule
\end{tabular*}
\end{minipage}
\end{minipage}
\end{table}

\begin{figure}[h]
    \centering
    \includegraphics[width=\linewidth]{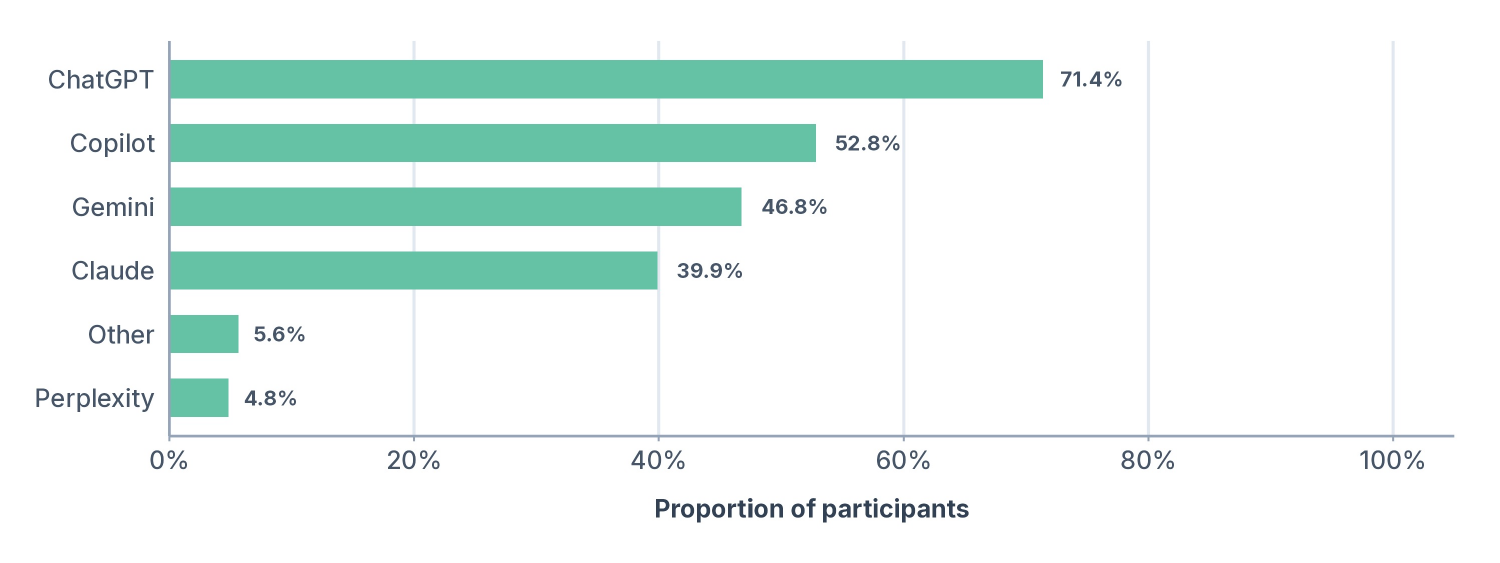}
    \caption{Proportion of participants ($n=248$) that reported using each AI tool. Participants could select multiple tools, so bars do not sum up to 100\%. The most commonly used AI tool was ChatGPT at 71.4\%.}
    \Description{TBD}
    \label{apx_fig:ai_tool_usage}
\end{figure}

\clearpage
\onecolumn
\section{Experimental Vignette Study}
\begin{table}[h!]
\centering
\caption{Vignette text stubs that are joined together to form a full vignette that participants then evaluate. \nonai{} vignettes combine \tasktype{} with corresponding text around how the task in question is completed without AI. \aienabled{} vignettes also start with \tasktype{} and are then formed by appending \specialization{} and \visibility{}, and ending with \role{}. Across the vignettes, Sam is one of four possible target helper colleagues. Sam's name is retained in the text stubs for readability, but the name shown to participants in the vignette is randomized by our custom online portal.}
\label{apx_tab:vignettes}
\scriptsize
\setlength{\tabcolsep}{3pt}
\renewcommand{\arraystretch}{1.05}
\begin{tabularx}{\linewidth}{
    @{}
    l
    l
    l
    l|
    >{\raggedright\arraybackslash}X
    @{}
}
\multicolumn{2}{l}{} &
\multicolumn{2}{l|}{} &
\textbf{Vignette Text} \\ \Xhline{1pt}
\multicolumn{2}{l|}{} & \multicolumn{2}{l|}{\assurance{}} & You have been assigned to prepare a section of the presentation summarizing lessons learned from a project that your team and peer colleagues at NextGen recently completed. As you work on the section, you begin to second-guess whether your perspective on what happened during the project is reasonable. Sam is working on another section of the same presentation. \\ \cline{3-5} 
\multicolumn{2}{l|}{} & \multicolumn{2}{l|}{\brainstorming{}} & You have been assigned to prepare a section of the presentation proposing ideas for new products or features that NextGen could develop in response to recent industry trends. Sam is working on another section of the same presentation. \\ \cline{3-5} 
\multicolumn{2}{l|}{} & \multicolumn{2}{l|}{\proofing{}} & You have been assigned to prepare a section of the presentation explaining the implications of recent industry trends for NextGen. You have already prepared an initial draft of your section, but think there may be ways to improve it. Sam is working on another section of the same presentation. \\ \cline{3-5} 
\multicolumn{2}{l|}{\multirow{-9}{*}{\tasktype{}}} & \multicolumn{2}{l|}{\search{}} & You have been assigned to prepare a section of the presentation where you need to search for recent products launched by NextGen's competitors. Sam is working on another section of the same presentation. \\ \Xhline{1pt}
\multicolumn{3}{l|}{} & \assurance{} & For this task, you revisit the project materials and think through which takeaways are most important. You use this review to consider which takeaways to include and continue preparing your section of the presentation. \\ \cline{4-5} 
\multicolumn{3}{l|}{} & \brainstorming{} & For this task, you brainstorm possible ideas for new products or features that NextGen could develop. You consider the ideas you generate, consider which ones to develop, and continue preparing your section of the presentation. \\ \cline{4-5} 
\multicolumn{3}{l|}{} & \proofing{} & For this task, you review your initial draft, considering its clarity, organization, and completeness. You identify potential revisions based on your review and continue preparing your section of the presentation. \\ \cline{4-5} 
\multicolumn{3}{l|}{\multirow{-7}{*}{\nonai{}}} & \search{} & For this task, you search company websites and other publicly available sources to identify recent products launched by competing companies and other relevant competitors to examine. You review the information you find and use it to continue preparing your section of the presentation. \\ \Xhline{1pt}
\multicolumn{1}{l|}{} & \multicolumn{1}{l|}{} & \multicolumn{1}{l|}{} & \assurance{} & You tell {[}OpenAssist/CorpAssist{]} that you are second-guessing your own perspective and ask it for help to work through your uncertainty. It offers considerations and alternative perspectives to help you think it through. You reflect on its response and decide for yourself whether your interpretation is reasonable. \\ \cline{4-5} 
\multicolumn{1}{l|}{} & \multicolumn{1}{l|}{} & \multicolumn{1}{l|}{} & \brainstorming{} & {\color[HTML]{000000} You ask {[}OpenAssist/CorpAssist{]} to generate possible ideas for new products or features that NextGen could develop. You use its suggestions as a starting point, decide which ideas to develop, and prepare your section of the presentation yourself.} \\ \cline{4-5} 
\multicolumn{1}{l|}{} & \multicolumn{1}{l|}{} & \multicolumn{1}{l|}{} & \proofing{} & {\color[HTML]{000000} You ask {[}OpenAssist/CorpAssist{]} to help review your draft and provide feedback on its clarity, organization, and completeness. You review its feedback, decide which changes to make, and revise your section of the presentation yourself accordingly.} \\ \cline{4-5} 
\multicolumn{1}{l|}{} & \multicolumn{1}{l|}{} & \multicolumn{1}{l|}{\multirow{-9}{*}{\assisting{}}} & \search{} & {\color[HTML]{000000} You ask {[}OpenAssist/CorpAssist{]} to identify recent products launched by competing companies and other relevant competitors of NextGen's to examine. You prepare your section of the presentation yourself, using its output to support your preparation.} \\ \cline{3-5} 
\multicolumn{1}{l|}{} & \multicolumn{1}{l|}{} & \multicolumn{1}{l|}{} & \assurance{} & {\color[HTML]{000000} You tell {[}OpenAssist/CorpAssist{]} that you are second-guessing your own perspective and ask it to assess your interpretation and determine whether your perspective is reasonable. You rely on its assessment as you continue preparing your section.} \\ \cline{4-5} 
\multicolumn{1}{l|}{} & \multicolumn{1}{l|}{} & \multicolumn{1}{l|}{} & \brainstorming{} & {\color[HTML]{000000} You ask {[}OpenAssist/CorpAssist{]} to prepare your section of the presentation for you by researching recent industry trends, generating ideas, and developing them into proposals for new products or features for NextGen.} \\ \cline{4-5} 
\multicolumn{1}{l|}{} & \multicolumn{1}{l|}{} & \multicolumn{1}{l|}{} & \proofing{} & {\color[HTML]{000000} You ask {[}OpenAssist/CorpAssist{]} to revise your draft for you to improve its clarity, organization, and completeness.} \\ \cline{4-5} 
\multicolumn{1}{l|}{} & \multicolumn{1}{l|}{\multirow{-24}{*}{\role{}}} & \multicolumn{1}{l|}{\multirow{-10}{*}{\delegating{}}} & \search{} & You ask {[}OpenAssist/CorpAssist{]} to prepare your section of the presentation for you by identifying and synthesizing recent products by NextGen's competitors. \\ \cline{2-5} 
\multicolumn{1}{l|}{} & \multicolumn{1}{l|}{} & \multicolumn{2}{l|}{\openassist{}} & {\color[HTML]{000000} For this task, you use OpenAssist, the general-purpose version of Assist, ...} \\ \cline{3-5} 
\multicolumn{1}{l|}{} & \multicolumn{1}{l|}{\multirow{-2}{*}{\specialization{}}} & \multicolumn{2}{l|}{\corpassist{}} & {\color[HTML]{000000} For this task, you use CorpAssist, the company-specific version of Assist, ...} \\ \cline{2-5} 
\multicolumn{1}{l|}{} & \multicolumn{1}{l|}{} & \multicolumn{2}{l|}{\personal} & ...individually on your own computer. \\ \cline{3-5} 
\multicolumn{1}{l|}{\multirow{-30}{*}{\aienabled{}}} & \multicolumn{1}{l|}{\multirow{-2}{*}{\visibility{}}} & \multicolumn{2}{l|}{\shared} & ...in the team’s shared workspace. \\ 
\end{tabularx}
\end{table}

\begin{figure}[h]
    \centering
    \includegraphics[width=0.9\linewidth]{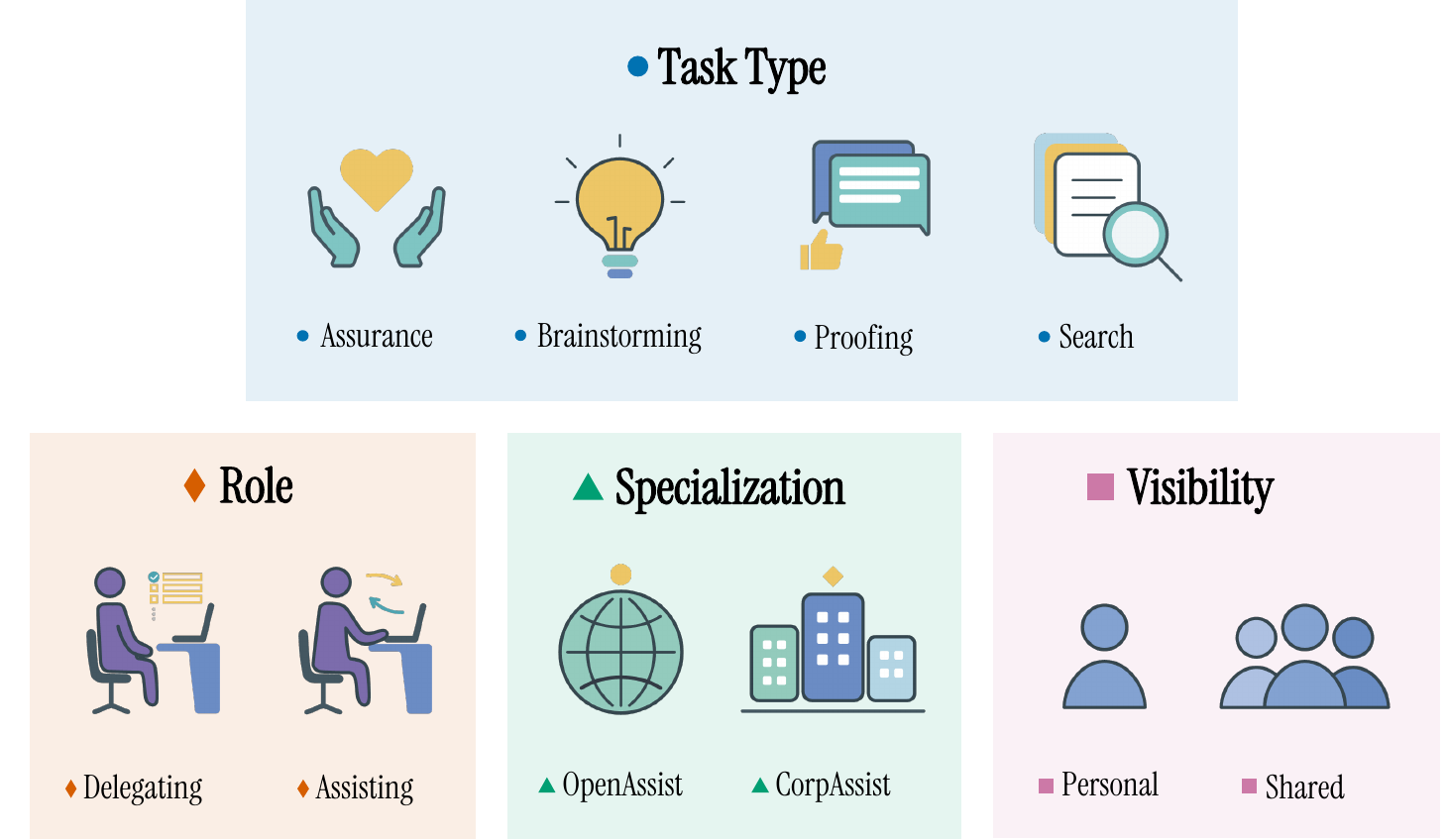}
    \caption{The full set of visual icons we designed to represent each factor level. Icons were presented alongside the vignette to facilitate comprehension.}
    \Description{TBD}
    \label{apx_fig:factor_icons}
\end{figure}

\begin{figure}[h]
    \centering
    \includegraphics[width=0.7\linewidth]{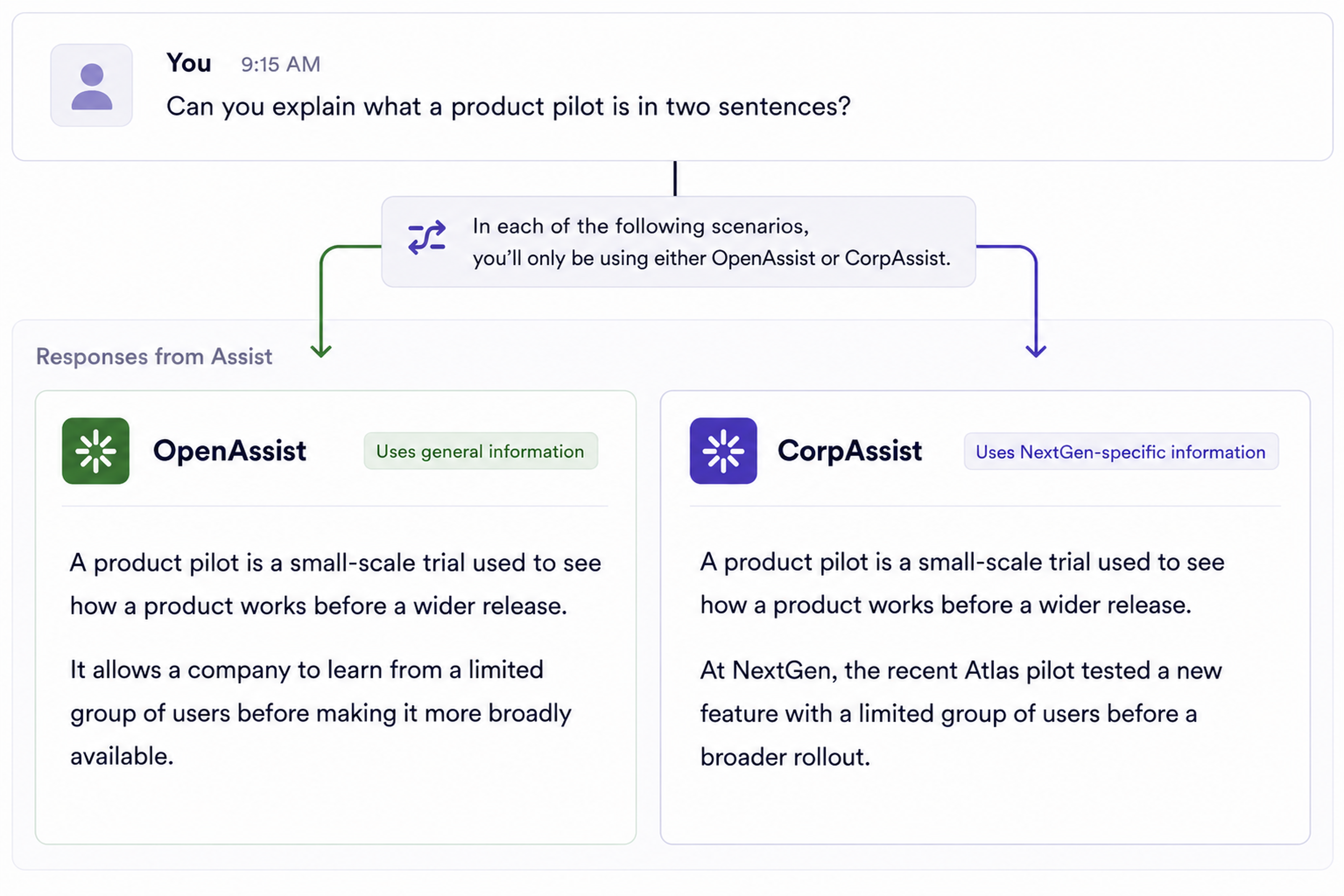}
    \caption{Diagram that provides an example of the output one might expect to receive from \openassist{} compared to \corpassist{}, displayed in the orientation of the custom online vignette portal to help participants differentiate between the two conditions.}
    \Description{TBD}
    \label{apx_fig:specialization_bg}
\end{figure}

\begin{figure}[h]
    \centering
    \includegraphics[width=\linewidth]{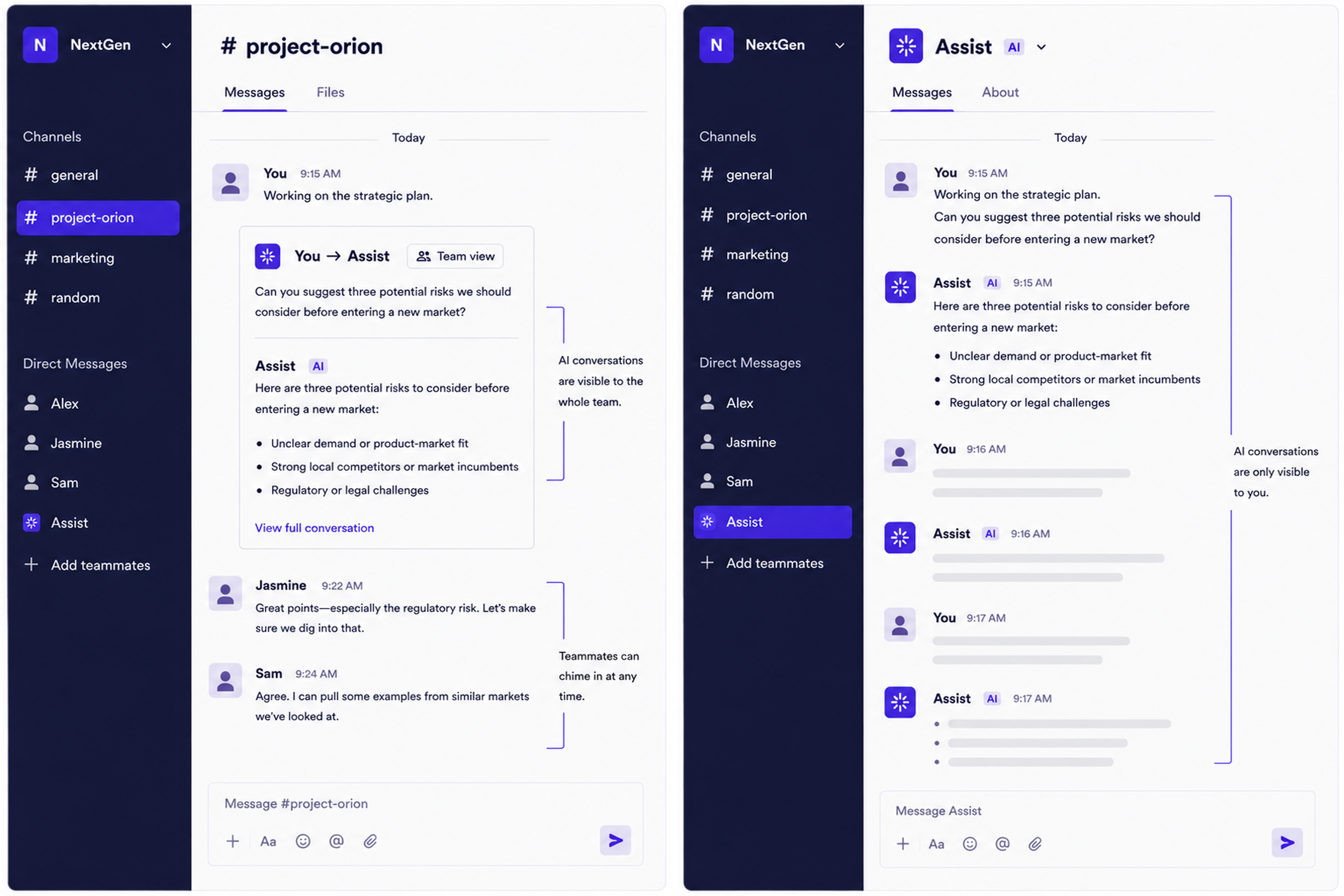}
    \caption{Side-by-side example screenshots of what \shared{} versus \personal{} AI usage might look like, situated in a generic workplace messaging platform to bring these conditions to life. Displayed in the orientation of the custom online vignette portal.}
    \Description{TBD}
    \label{apx_fig:visibility_bg}
\end{figure}

\clearpage
\onecolumn
\section{Equations of the Full Linear Mixed-Effects Models}
\label{apx_sec:models}

\begin{equation}
\label{eqn:rq1_full_model}
\begin{split}
    Y_{d} \sim{} & ai\_presence + task\_type + ai\_presence:task\_type + order\_vig \\
    & + gender + age + job\_function + job\_role + tenure + num\_direct\_reports \\
    & + work\_arrangement + collab\_frequency + num\_colleagues\_interact \\
    & + org\_size + org\_ai\_training + org\_ai\_encouragement \\
    & + tps\_score + social\_anxiety \\
    & + bfi\_extraversion + bfi\_agreeableness + bfi\_neuroticism \\
    & + bfi\_conscientiousness + bfi\_openness \\
    & + feedback\_utility + feedback\_accountability \\
    & + feedback\_social\_awareness + feedback\_self\_efficacy \\
    & + (1 | participant) \\
    d \in{} & \{\mathrm{\dvhs{}}, \mathrm{\dvcc{}}\}
\end{split}
\end{equation}

\begin{equation}
\label{eqn:rq2_full_model}
\begin{split}
    Y_{d} \sim{} & task\_type + role + specialization + visibility + order\_vig \\
    & + gender + age + job\_function + job\_role + tenure \\
    & + num\_direct\_reports + work\_arrangement + collab\_frequency \\
    & + num\_colleagues\_interact + org\_size + org\_ai\_training \\
    & + org\_ai\_encouragement + tps\_score + social\_anxiety \\
    & + bfi\_extraversion + bfi\_agreeableness + bfi\_neuroticism \\
    & + bfi\_conscientiousness + bfi\_openness \\
    & + feedback\_utility + feedback\_accountability \\
    & + feedback\_social\_awareness + feedback\_self\_efficacy \\
    & + (1 | participant) \\
    d \in{} & \{\mathrm{\dvhs{}}, \mathrm{\dvcc{}}\}
\end{split}
\end{equation}

\clearpage
\onecolumn
\section{Full Results of Linear Mixed-Effects Models}
\label{apx_sec:model_tables}

\definecolor{editCol}{RGB}{0,0,0}
\definecolor{lightline}{RGB}{210,210,210}
\begin{table*}[th!]
\centering
\scriptsize
\setlength{\tabcolsep}{6pt}
\caption{\textcolor{editCol}{\textbf{Incrementally built models for RQ1}: estimated effects from Linear Mixed-Effects Models for \dvhs{} ($M_{RQ1,HS}$). $M_{RQ1,HS,a}$ includes the most basic model. Intermediate models are provided, where $M_{RQ1,HS,b}$ adds demographics and work context; $M_{RQ1,HS,c}$ adds organizational context; and $M_{RQ1,HS}$ adds individual personality to form the full model. Table entries report coefficient estimates and corresponding $p$-values ($\cdot\ p < 0.10$, * $p < 0.05$, ** $p < 0.01$, *** $p < 0.001$).}}
\begin{tabular}{>{\color{editCol}}r|>{\color{editCol}}l>{\color{editCol}}l>{\color{editCol}}l|>{\color{editCol}}l>{\color{editCol}}l>{\color{editCol}}l|>{\color{editCol}}l>{\color{editCol}}l>{\color{editCol}}l|>{\color{editCol}}l>{\color{editCol}}l>{\color{editCol}}l}
& \multicolumn{3}{>{\color{editCol}}c|}{$M_{RQ1,HS,a}$} & \multicolumn{3}{>{\color{editCol}}c|}{$M_{RQ1,HS,b}$} & \multicolumn{3}{>{\color{editCol}}c|}{$M_{RQ1,HS,c}$} & \multicolumn{3}{>{\color{editCol}}c}{$M_{RQ1,HS}$} \\
& Est. & \multicolumn{2}{>{\color{editCol}}c|}{$p$-value} & Est. & \multicolumn{2}{>{\color{editCol}}c|}{$p$-value} & Est. & \multicolumn{2}{>{\color{editCol}}c|}{$p$-value} & Est. & \multicolumn{2}{>{\color{editCol}}c}{$p$-value} \\
\hline
\arrayrulecolor{lightline}
\multicolumn{13}{>{\color{editCol}}l}{\textbf{H1: \aipresence{} (ref: \aienabled{})}} \\
\quad \nonai{} & $0.163$ & $0.016$ & * & $0.165$ & $0.015$ & * & $0.164$ & $0.015$ & * & $0.166$ & $0.014$ & * \\
\hline
\multicolumn{13}{>{\color{editCol}}l}{\textbf{H2: \tasktype[Task Type] (ref: \assurance{})}} \\
\quad \brainstorming{} & $-0.295$ & $< .001$ & *** & $-0.295$ & $< .001$ & *** & $-0.295$ & $< .001$ & *** & $-0.295$ & $< .001$ & *** \\
\quad \proofing{} & $-0.234$ & $< .001$ & *** & $-0.234$ & $< .001$ & *** & $-0.234$ & $< .001$ & *** & $-0.234$ & $< .001$ & *** \\
\quad \search{} & $-0.327$ & $< .001$ & *** & $-0.327$ & $< .001$ & *** & $-0.327$ & $< .001$ & *** & $-0.327$ & $< .001$ & *** \\
\hline
\multicolumn{13}{>{\color{editCol}}l}{\textbf{\aipresence{} $\times$ \tasktype[Task Type] (ref: \aienabled{} $\times$ \assurance{})}} \\
\nonai{} $\times$ \brainstorming{} & $0.247$ & $0.010$ & ** & $0.246$ & $0.010$ & * & $0.248$ & $0.010$ & ** & $0.247$ & $0.010$ & ** \\
\nonai{} $\times$ \proofing{} & $0.100$ & $0.285$ &  & $0.096$ & $0.306$ &  & $0.096$ & $0.304$ &  & $0.095$ & $0.309$ &  \\
\nonai{} $\times$ \search{} & $-0.040$ & $0.670$ &  & $-0.042$ & $0.652$ &  & $-0.041$ & $0.659$ &  & $-0.044$ & $0.638$ &  \\
\hline
\multicolumn{13}{>{\color{editCol}}l}{\textbf{Vignette Order}} \\
Vignette Number & $-0.009$ & $0.066$ & $\cdot$ & $-0.009$ & $0.070$ & $\cdot$ & $-0.009$ & $0.067$ & $\cdot$ & $-0.009$ & $0.068$ & $\cdot$ \\
\hline
\multicolumn{13}{>{\color{editCol}}l}{\textbf{Demographics and Work}} \\
\multicolumn{13}{>{\color{editCol}}l}{\textit{Gender (ref: Man)}} \\
\quad Woman &  &  &  & $0.004$ & $0.955$ &  & $-0.027$ & $0.721$ &  & $-0.022$ & $0.772$ &  \\
\quad Other &  &  &  & $0.089$ & $0.753$ &  & $-0.005$ & $0.987$ &  & $-0.137$ & $0.627$ &  \\
\multicolumn{13}{>{\color{editCol}}l}{\textit{Age (ref: 18--24)}} \\
\quad 25--34 &  &  &  & $-0.119$ & $0.356$ &  & $-0.034$ & $0.797$ &  & $-0.041$ & $0.754$ &  \\
\quad 35--44 &  &  &  & $-0.104$ & $0.478$ &  & $-0.033$ & $0.823$ &  & $-0.036$ & $0.806$ &  \\
\quad 45--54 &  &  &  & $-0.303$ & $0.053$ & $\cdot$ & $-0.243$ & $0.123$ &  & $-0.186$ & $0.240$ &  \\
\quad 55--64 &  &  &  & $-0.397$ & $0.052$ & $\cdot$ & $-0.319$ & $0.118$ &  & $-0.303$ & $0.134$ &  \\
\multicolumn{13}{>{\color{editCol}}l}{\textit{Job Function (ref: Analysis / Research)}} \\
\quad Software / Engineering &  &  &  & $0.026$ & $0.794$ &  & $0.014$ & $0.891$ &  & $0.054$ & $0.596$ &  \\
\quad Planning / Project Management &  &  &  & $-0.142$ & $0.182$ &  & $-0.126$ & $0.238$ &  & $-0.096$ & $0.367$ &  \\
\quad Writing / Content Creation &  &  &  & $-0.091$ & $0.447$ &  & $-0.063$ & $0.603$ &  & $-0.029$ & $0.809$ &  \\
\quad Management / Strategy &  &  &  & $-0.101$ & $0.490$ &  & $-0.101$ & $0.491$ &  & $-0.067$ & $0.647$ &  \\
\quad Design / Creative Work &  &  &  & $0.084$ & $0.571$ &  & $0.096$ & $0.538$ &  & $0.167$ & $0.275$ &  \\
\multicolumn{13}{>{\color{editCol}}l}{\textit{Job Role (ref: Managers)}} \\
\quad Skilled Office Worker &  &  &  & $-0.372$ & $0.010$ & ** & $-0.401$ & $0.006$ & ** & $-0.461$ & $0.002$ & ** \\
\quad Senior Staff &  &  &  & $-0.313$ & $0.029$ & * & $-0.337$ & $0.022$ & * & $-0.393$ & $0.008$ & ** \\
\multicolumn{13}{>{\color{editCol}}l}{\textit{Tenure (ref: $\leq$2 years)}} \\
\quad 3--5 years &  &  &  & $-0.055$ & $0.539$ &  & $-0.085$ & $0.344$ &  & $-0.063$ & $0.488$ &  \\
\quad 6+ years &  &  &  & $0.074$ & $0.454$ &  & $0.021$ & $0.836$ &  & $-0.002$ & $0.988$ &  \\
\multicolumn{13}{>{\color{editCol}}l}{\textit{Work Arrangement (ref: In-Person)}} \\
\quad Remote &  &  &  & $0.012$ & $0.897$ &  & $-0.003$ & $0.976$ &  & $-0.012$ & $0.895$ &  \\
\quad Hybrid &  &  &  & $-0.010$ & $0.912$ &  & $-0.033$ & $0.720$ &  & $-0.030$ & $0.739$ &  \\
Has Direct Reports &  &  &  & $-0.134$ & $0.253$ &  & $-0.115$ & $0.328$ &  & $-0.149$ & $0.200$ &  \\
\hline
\multicolumn{13}{>{\color{editCol}}l}{\textbf{Organizational Context}} \\
Collaboration Frequency &  &  &  & $0.058$ & $0.261$ &  & $0.054$ & $0.304$ &  & $-0.005$ & $0.919$ &  \\
Number of Colleagues &  &  &  & $-0.007$ & $0.891$ &  & $-0.041$ & $0.415$ &  & $-0.042$ & $0.406$ &  \\
Org. Has AI Training &  &  &  &  &  &  & $-0.009$ & $0.784$ &  & $-0.036$ & $0.298$ &  \\
Org. Encourages AI &  &  &  &  &  &  & $-0.009$ & $0.863$ &  & $-0.030$ & $0.544$ &  \\
\multicolumn{13}{>{\color{editCol}}l}{\textit{Organization Size (ref: <50)}} \\
\quad 50-249 &  &  &  &  &  &  & $0.081$ & $0.570$ &  & $0.043$ & $0.762$ &  \\
\quad 250-999 &  &  &  &  &  &  & $-0.144$ & $0.334$ &  & $-0.151$ & $0.314$ &  \\
\quad 1,000-9,999 &  &  &  &  &  &  & $0.002$ & $0.986$ &  & $0.029$ & $0.830$ &  \\
\quad 10,000 or more &  &  &  &  &  &  & $0.203$ & $0.162$ &  & $0.252$ & $0.081$ & $\cdot$ \\
\hline
\multicolumn{13}{>{\color{editCol}}l}{\textbf{Individual Personality}} \\
Team Psychological Safety &  &  &  &  &  &  &  &  &  & $0.022$ & $0.574$ &  \\
\multicolumn{13}{>{\color{editCol}}l}{\textit{Big-5 Personality}} \\
Openness &  &  &  &  &  &  &  &  &  & $0.014$ & $0.696$ &  \\
Conscientiousness &  &  &  &  &  &  &  &  &  & $0.005$ & $0.916$ &  \\
Extraversion &  &  &  &  &  &  &  &  &  & $-0.001$ & $0.979$ &  \\
Agreeableness &  &  &  &  &  &  &  &  &  & $0.053$ & $0.184$ &  \\
Neuroticism &  &  &  &  &  &  &  &  &  & $0.012$ & $0.760$ &  \\
\multicolumn{13}{>{\color{editCol}}l}{\textit{Feedback Orientation Scale}} \\
Feedback Utility &  &  &  &  &  &  &  &  &  & $0.169$ & $0.015$ & * \\
Feedback Accountability &  &  &  &  &  &  &  &  &  & $-0.025$ & $0.755$ &  \\
Feedback Social Awareness &  &  &  &  &  &  &  &  &  & $0.083$ & $0.124$ &  \\
Feedback Self-Efficacy &  &  &  &  &  &  &  &  &  & $0.054$ & $0.359$ &  \\
\multicolumn{13}{>{\color{editCol}}l}{\textit{Social Anxiety (ref: High)}} \\
\quad Low &  &  &  &  &  &  &  &  &  & $-0.132$ & $0.110$ &  \\
\hline
\arrayrulecolor{black}
Conditional $R^2$ & \multicolumn{3}{>{\color{editCol}}c|}{0.399} & \multicolumn{3}{>{\color{editCol}}c|}{0.418} & \multicolumn{3}{>{\color{editCol}}c|}{0.424} & \multicolumn{3}{>{\color{editCol}}c}{0.431} \\
\hline
\end{tabular}
\label{apx_tab:rq1_full_hs}
\end{table*}

\definecolor{editCol}{RGB}{0,0,0}
\definecolor{lightline}{RGB}{210,210,210}
\begin{table*}[th!]
\centering
\scriptsize
\setlength{\tabcolsep}{6pt}
\caption{\textcolor{editCol}{\textbf{Incrementally built models for RQ1}: estimated effects from Linear Mixed-Effects Models for \dvcc{} ($M_{RQ1,CC}$). $M_{RQ1,CC,a}$ includes the most basic model. Intermediate models are provided, where $M_{RQ1,CC,b}$ adds demographics and work context; $M_{RQ1,CC,c}$ adds organizational context; and $M_{RQ1,CC}$ adds individual personality to form the full model. Table entries report coefficient estimates and corresponding $p$-values ($\cdot\ p < 0.10$, * $p < 0.05$, ** $p < 0.01$, *** $p < 0.001$).}}
\begin{tabular}{>{\color{editCol}}r|>{\color{editCol}}l>{\color{editCol}}l>{\color{editCol}}l|>{\color{editCol}}l>{\color{editCol}}l>{\color{editCol}}l|>{\color{editCol}}l>{\color{editCol}}l>{\color{editCol}}l|>{\color{editCol}}l>{\color{editCol}}l>{\color{editCol}}l}
& \multicolumn{3}{>{\color{editCol}}c|}{$M_{RQ1,CC,a}$} & \multicolumn{3}{>{\color{editCol}}c|}{$M_{RQ1,CC,b}$} & \multicolumn{3}{>{\color{editCol}}c|}{$M_{RQ1,CC,c}$} & \multicolumn{3}{>{\color{editCol}}c}{$M_{RQ1,CC}$} \\
& Est. & \multicolumn{2}{>{\color{editCol}}c|}{$p$-value} & Est. & \multicolumn{2}{>{\color{editCol}}c|}{$p$-value} & Est. & \multicolumn{2}{>{\color{editCol}}c|}{$p$-value} & Est. & \multicolumn{2}{>{\color{editCol}}c}{$p$-value} \\
\hline
\arrayrulecolor{lightline}
\multicolumn{13}{>{\color{editCol}}l}{\textbf{H1: \aipresence{} (ref: \aienabled{})}} \\
\quad \nonai{} & $0.229$ & $0.003$ & ** & $0.232$ & $0.003$ & ** & $0.231$ & $0.003$ & ** & $0.233$ & $0.003$ & ** \\
\hline
\multicolumn{13}{>{\color{editCol}}l}{\textbf{H2: \tasktype[Task Type] (ref: \assurance{})}} \\
\quad \brainstorming{} & $-0.285$ & $< .001$ & *** & $-0.285$ & $< .001$ & *** & $-0.285$ & $< .001$ & *** & $-0.285$ & $< .001$ & *** \\
\quad \proofing{} & $-0.236$ & $< .001$ & *** & $-0.236$ & $< .001$ & *** & $-0.236$ & $< .001$ & *** & $-0.237$ & $< .001$ & *** \\
\quad \search{} & $-0.330$ & $< .001$ & *** & $-0.330$ & $< .001$ & *** & $-0.330$ & $< .001$ & *** & $-0.330$ & $< .001$ & *** \\
\hline
\multicolumn{13}{>{\color{editCol}}l}{\textbf{\aipresence{} $\times$ \tasktype[Task Type] (ref: \aienabled{} $\times$ \assurance{})}} \\
\nonai{} $\times$ \brainstorming{} & $0.157$ & $0.156$ &  & $0.156$ & $0.158$ &  & $0.156$ & $0.158$ &  & $0.155$ & $0.159$ &  \\
\nonai{} $\times$ \proofing{} & $-0.040$ & $0.709$ &  & $-0.045$ & $0.677$ &  & $-0.045$ & $0.675$ &  & $-0.048$ & $0.657$ &  \\
\nonai{} $\times$ \search{} & $-0.047$ & $0.665$ &  & $-0.048$ & $0.657$ &  & $-0.048$ & $0.658$ &  & $-0.050$ & $0.646$ &  \\
\hline
\multicolumn{13}{>{\color{editCol}}l}{\textbf{Vignette Order}} \\
Vignette Number & $-0.007$ & $0.204$ &  & $-0.007$ & $0.214$ &  & $-0.007$ & $0.205$ &  & $-0.007$ & $0.208$ &  \\
\hline
\multicolumn{13}{>{\color{editCol}}l}{\textbf{Demographics and Work}} \\
\multicolumn{13}{>{\color{editCol}}l}{\textit{Gender (ref: Man)}} \\
\quad Woman &  &  &  & $0.025$ & $0.758$ &  & $-0.006$ & $0.940$ &  & $0.017$ & $0.836$ &  \\
\quad Other &  &  &  & $0.191$ & $0.532$ &  & $0.097$ & $0.753$ &  & $0.034$ & $0.912$ &  \\
\multicolumn{13}{>{\color{editCol}}l}{\textit{Age (ref: 18--24)}} \\
\quad 25--34 &  &  &  & $-0.212$ & $0.130$ &  & $-0.136$ & $0.339$ &  & $-0.143$ & $0.321$ &  \\
\quad 35--44 &  &  &  & $-0.103$ & $0.518$ &  & $-0.023$ & $0.886$ &  & $-0.016$ & $0.921$ &  \\
\quad 45--54 &  &  &  & $-0.381$ & $0.025$ & * & $-0.340$ & $0.046$ & * & $-0.308$ & $0.078$ & $\cdot$ \\
\quad 55--64 &  &  &  & $-0.449$ & $0.042$ & * & $-0.383$ & $0.082$ & $\cdot$ & $-0.388$ & $0.083$ & $\cdot$ \\
\multicolumn{13}{>{\color{editCol}}l}{\textit{Job Function (ref: Analysis / Research)}} \\
\quad Software / Engineering &  &  &  & $-0.014$ & $0.894$ &  & $-0.038$ & $0.726$ &  & $-0.019$ & $0.869$ &  \\
\quad Planning / Project Management &  &  &  & $-0.223$ & $0.054$ & $\cdot$ & $-0.229$ & $0.048$ & * & $-0.211$ & $0.073$ & $\cdot$ \\
\quad Writing / Content Creation &  &  &  & $-0.039$ & $0.761$ &  & $0.014$ & $0.911$ &  & $0.057$ & $0.670$ &  \\
\quad Management / Strategy &  &  &  & $0.031$ & $0.843$ &  & $0.036$ & $0.818$ &  & $0.037$ & $0.818$ &  \\
\quad Design / Creative Work &  &  &  & $0.041$ & $0.798$ &  & $0.100$ & $0.549$ &  & $0.152$ & $0.370$ &  \\
\multicolumn{13}{>{\color{editCol}}l}{\textit{Job Role (ref: Managers)}} \\
\quad Skilled Office Worker &  &  &  & $-0.307$ & $0.049$ & * & $-0.315$ & $0.045$ & * & $-0.336$ & $0.036$ & * \\
\quad Senior Staff &  &  &  & $-0.304$ & $0.051$ & $\cdot$ & $-0.320$ & $0.043$ & * & $-0.340$ & $0.036$ & * \\
\multicolumn{13}{>{\color{editCol}}l}{\textit{Tenure (ref: $\leq$2 years)}} \\
\quad 3--5 years &  &  &  & $0.025$ & $0.798$ &  & $-0.024$ & $0.806$ &  & $0.006$ & $0.954$ &  \\
\quad 6+ years &  &  &  & $0.168$ & $0.119$ &  & $0.103$ & $0.341$ &  & $0.084$ & $0.453$ &  \\
\multicolumn{13}{>{\color{editCol}}l}{\textit{Work Arrangement (ref: In-Person)}} \\
\quad Remote &  &  &  & $0.013$ & $0.897$ &  & $-0.008$ & $0.941$ &  & $0.000$ & $0.996$ &  \\
\quad Hybrid &  &  &  & $0.026$ & $0.796$ &  & $-0.004$ & $0.971$ &  & $0.017$ & $0.863$ &  \\
Has Direct Reports &  &  &  & $-0.099$ & $0.436$ &  & $-0.065$ & $0.610$ &  & $-0.076$ & $0.553$ &  \\
\hline
\multicolumn{13}{>{\color{editCol}}l}{\textbf{Organizational Context}} \\
Collaboration Frequency &  &  &  & $0.027$ & $0.638$ &  & $0.019$ & $0.741$ &  & $-0.024$ & $0.689$ &  \\
Number of Colleagues &  &  &  & $0.031$ & $0.564$ &  & $-0.006$ & $0.912$ &  & $0.006$ & $0.911$ &  \\
Org. Has AI Training &  &  &  &  &  &  & $-0.044$ & $0.221$ &  & $-0.070$ & $0.068$ & $\cdot$ \\
Org. Encourages AI &  &  &  &  &  &  & $0.058$ & $0.279$ &  & $0.039$ & $0.473$ &  \\
\multicolumn{13}{>{\color{editCol}}l}{\textit{Organization Size (ref: <50)}} \\
\quad 50-249 &  &  &  &  &  &  & $-0.032$ & $0.837$ &  & $-0.077$ & $0.625$ &  \\
\quad 250-999 &  &  &  &  &  &  & $-0.110$ & $0.494$ &  & $-0.146$ & $0.379$ &  \\
\quad 1,000-9,999 &  &  &  &  &  &  & $0.063$ & $0.668$ &  & $0.055$ & $0.711$ &  \\
\quad 10,000 or more &  &  &  &  &  &  & $0.249$ & $0.113$ &  & $0.262$ & $0.100$ &  \\
\hline
\multicolumn{13}{>{\color{editCol}}l}{\textbf{Individual Personality}} \\
Team Psychological Safety &  &  &  &  &  &  &  &  &  & $0.054$ & $0.218$ &  \\
\multicolumn{13}{>{\color{editCol}}l}{\textit{Big-5 Personality}} \\
Openness &  &  &  &  &  &  &  &  &  & $0.007$ & $0.849$ &  \\
Conscientiousness &  &  &  &  &  &  &  &  &  & $0.018$ & $0.739$ &  \\
Extraversion &  &  &  &  &  &  &  &  &  & $-0.009$ & $0.826$ &  \\
Agreeableness &  &  &  &  &  &  &  &  &  & $0.037$ & $0.398$ &  \\
Neuroticism &  &  &  &  &  &  &  &  &  & $-0.014$ & $0.747$ &  \\
\multicolumn{13}{>{\color{editCol}}l}{\textit{Feedback Orientation Scale}} \\
Feedback Utility &  &  &  &  &  &  &  &  &  & $0.134$ & $0.081$ & $\cdot$ \\
Feedback Accountability &  &  &  &  &  &  &  &  &  & $-0.080$ & $0.373$ &  \\
Feedback Social Awareness &  &  &  &  &  &  &  &  &  & $0.074$ & $0.215$ &  \\
Feedback Self-Efficacy &  &  &  &  &  &  &  &  &  & $0.037$ & $0.565$ &  \\
\multicolumn{13}{>{\color{editCol}}l}{\textit{Social Anxiety (ref: High)}} \\
\quad Low &  &  &  &  &  &  &  &  &  & $-0.052$ & $0.569$ &  \\
\hline
\arrayrulecolor{black}
Conditional $R^2$ & \multicolumn{3}{>{\color{editCol}}c|}{0.362} & \multicolumn{3}{>{\color{editCol}}c|}{0.381} & \multicolumn{3}{>{\color{editCol}}c|}{0.386} & \multicolumn{3}{>{\color{editCol}}c}{0.395} \\
\hline
\end{tabular}
\label{apx_tab:rq1_full_cc}
\end{table*}

\definecolor{editCol}{RGB}{0,0,0}
\definecolor{lightline}{RGB}{210,210,210}
\begin{table*}[th!]
\centering
\scriptsize
\setlength{\tabcolsep}{6pt}
\caption{\textcolor{editCol}{\textbf{Incrementally built models for RQ2}: estimated effects from Linear Mixed-Effects Models for \dvhs{} ($M_{RQ2,HS}$). $M_{RQ2,HS,a}$ includes the most basic model. Intermediate models are provided, where $M_{RQ2,HS,b}$ adds demographics and work context; $M_{RQ2,HS,c}$ adds organizational context; and $M_{RQ2,HS}$ adds individual personality to form the full model. Table entries report coefficient estimates and corresponding $p$-values ($\cdot\ p < 0.10$, * $p < 0.05$, ** $p < 0.01$, *** $p < 0.001$).}}
\begin{tabular}{>{\color{editCol}}r|>{\color{editCol}}l>{\color{editCol}}l>{\color{editCol}}l|>{\color{editCol}}l>{\color{editCol}}l>{\color{editCol}}l|>{\color{editCol}}l>{\color{editCol}}l>{\color{editCol}}l|>{\color{editCol}}l>{\color{editCol}}l>{\color{editCol}}l}
& \multicolumn{3}{>{\color{editCol}}c|}{$M_{RQ2,HS,a}$} & \multicolumn{3}{>{\color{editCol}}c|}{$M_{RQ2,HS,b}$} & \multicolumn{3}{>{\color{editCol}}c|}{$M_{RQ2,HS,c}$} & \multicolumn{3}{>{\color{editCol}}c}{$M_{RQ2,HS}$} \\
& Est. & \multicolumn{2}{>{\color{editCol}}c|}{$p$-value} & Est. & \multicolumn{2}{>{\color{editCol}}c|}{$p$-value} & Est. & \multicolumn{2}{>{\color{editCol}}c|}{$p$-value} & Est. & \multicolumn{2}{>{\color{editCol}}c}{$p$-value} \\
\hline
\arrayrulecolor{lightline}
\multicolumn{13}{>{\color{editCol}}l}{\textbf{H2: \tasktype[Task Type] (ref: \assurance{})}} \\
\quad \brainstorming{} & $-0.295$ & $< .001$ & *** & $-0.295$ & $< .001$ & *** & $-0.295$ & $< .001$ & *** & $-0.295$ & $< .001$ & *** \\
\quad \proofing{} & $-0.234$ & $< .001$ & *** & $-0.234$ & $< .001$ & *** & $-0.234$ & $< .001$ & *** & $-0.234$ & $< .001$ & *** \\
\quad \search{} & $-0.328$ & $< .001$ & *** & $-0.328$ & $< .001$ & *** & $-0.328$ & $< .001$ & *** & $-0.328$ & $< .001$ & *** \\
\hline
\multicolumn{13}{>{\color{editCol}}l}{\textbf{H3: \role[Role] (ref: \assisting{})}} \\
\quad \delegating{} & $0.036$ & $0.198$ &  & $0.036$ & $0.197$ &  & $0.036$ & $0.198$ &  & $0.036$ & $0.199$ &  \\
\hline
\multicolumn{13}{>{\color{editCol}}l}{\textbf{H4: \specialization[Specialization] (ref: \openassist{})}} \\
\quad \corpassist{} & $-0.026$ & $0.352$ &  & $-0.026$ & $0.351$ &  & $-0.026$ & $0.350$ &  & $-0.026$ & $0.353$ &  \\
\hline
\multicolumn{13}{>{\color{editCol}}l}{\textbf{H5: \visibility[Visibility] (ref: \shared{})}} \\
\quad \personal{} & $-0.306$ & $< .001$ & *** & $-0.307$ & $< .001$ & *** & $-0.307$ & $< .001$ & *** & $-0.306$ & $< .001$ & *** \\
\hline
\multicolumn{13}{>{\color{editCol}}l}{\textbf{Vignette Order}} \\
Vignette Number & $-0.012$ & $0.035$ & * & $-0.012$ & $0.037$ & * & $-0.012$ & $0.036$ & * & $-0.012$ & $0.035$ & * \\
\hline
\multicolumn{13}{>{\color{editCol}}l}{\textbf{Demographics and Work}} \\
\multicolumn{13}{>{\color{editCol}}l}{\textit{Gender (ref: Man)}} \\
\quad Woman &  &  &  & $-0.002$ & $0.979$ &  & $-0.033$ & $0.680$ &  & $-0.026$ & $0.749$ &  \\
\quad Other &  &  &  & $0.055$ & $0.856$ &  & $-0.034$ & $0.911$ &  & $-0.159$ & $0.599$ &  \\
Age (Ordinal Rank) &  &  &  & $-0.095$ & $0.023$ & * & $-0.091$ & $0.029$ & * & $-0.078$ & $0.067$ & $\cdot$ \\
\multicolumn{13}{>{\color{editCol}}l}{\textit{Job Function (ref: Analysis / Research)}} \\
\quad Software / Engineering &  &  &  & $0.016$ & $0.880$ &  & $-0.002$ & $0.983$ &  & $0.054$ & $0.623$ &  \\
\quad Planning / Project Management &  &  &  & $-0.120$ & $0.287$ &  & $-0.104$ & $0.363$ &  & $-0.060$ & $0.596$ &  \\
\quad Writing / Content Creation &  &  &  & $-0.086$ & $0.501$ &  & $-0.057$ & $0.658$ &  & $-0.029$ & $0.825$ &  \\
\quad Management / Strategy &  &  &  & $-0.104$ & $0.504$ &  & $-0.100$ & $0.520$ &  & $-0.068$ & $0.664$ &  \\
\quad Design / Creative Work &  &  &  & $0.121$ & $0.441$ &  & $0.144$ & $0.383$ &  & $0.218$ & $0.184$ &  \\
\multicolumn{13}{>{\color{editCol}}l}{\textit{Job Role (ref: Managers)}} \\
\quad Skilled Office Worker &  &  &  & $-0.408$ & $0.008$ & ** & $-0.434$ & $0.005$ & ** & $-0.496$ & $0.002$ & ** \\
\quad Senior Staff &  &  &  & $-0.330$ & $0.031$ & * & $-0.349$ & $0.026$ & * & $-0.411$ & $0.009$ & ** \\
\multicolumn{13}{>{\color{editCol}}l}{\textit{Tenure (ref: $\leq$2 years)}} \\
\quad 3--5 years &  &  &  & $-0.057$ & $0.531$ &  & $-0.074$ & $0.420$ &  & $-0.058$ & $0.532$ &  \\
\quad 6+ years &  &  &  & $0.061$ & $0.557$ &  & $0.012$ & $0.910$ &  & $-0.003$ & $0.974$ &  \\
\multicolumn{13}{>{\color{editCol}}l}{\textit{Work Arrangement (ref: In-Person)}} \\
\quad Remote &  &  &  & $-0.005$ & $0.960$ &  & $-0.018$ & $0.856$ &  & $-0.031$ & $0.761$ &  \\
\quad Hybrid &  &  &  & $-0.035$ & $0.717$ &  & $-0.058$ & $0.552$ &  & $-0.052$ & $0.589$ &  \\
Has Direct Reports &  &  &  & $-0.154$ & $0.219$ &  & $-0.143$ & $0.253$ &  & $-0.180$ & $0.149$ &  \\
\hline
\multicolumn{13}{>{\color{editCol}}l}{\textbf{Organizational Context}} \\
Collaboration Frequency &  &  &  & $0.050$ & $0.367$ &  & $0.047$ & $0.407$ &  & $-0.017$ & $0.771$ &  \\
Number of Colleagues &  &  &  & $-0.016$ & $0.765$ &  & $-0.050$ & $0.359$ &  & $-0.052$ & $0.340$ &  \\
Org. Has AI Training &  &  &  &  &  &  & $0.001$ & $0.981$ &  & $-0.024$ & $0.511$ &  \\
Org. Encourages AI &  &  &  &  &  &  & $-0.018$ & $0.734$ &  & $-0.035$ & $0.509$ &  \\
\multicolumn{13}{>{\color{editCol}}l}{\textit{Organization Size (ref: <50)}} \\
\quad 50-249 &  &  &  &  &  &  & $0.078$ & $0.603$ &  & $0.035$ & $0.818$ &  \\
\quad 250-999 &  &  &  &  &  &  & $-0.141$ & $0.364$ &  & $-0.144$ & $0.356$ &  \\
\quad 1,000-9,999 &  &  &  &  &  &  & $0.029$ & $0.839$ &  & $0.056$ & $0.696$ &  \\
\quad 10,000 or more &  &  &  &  &  &  & $0.210$ & $0.169$ &  & $0.260$ & $0.087$ & $\cdot$ \\
\hline
\multicolumn{13}{>{\color{editCol}}l}{\textbf{Individual Personality}} \\
Team Psychological Safety &  &  &  &  &  &  &  &  &  & $0.018$ & $0.661$ &  \\
\multicolumn{13}{>{\color{editCol}}l}{\textit{Big-5 Personality}} \\
Openness &  &  &  &  &  &  &  &  &  & $0.012$ & $0.753$ &  \\
Conscientiousness &  &  &  &  &  &  &  &  &  & $0.013$ & $0.808$ &  \\
Extraversion &  &  &  &  &  &  &  &  &  & $-0.002$ & $0.968$ &  \\
Agreeableness &  &  &  &  &  &  &  &  &  & $0.047$ & $0.266$ &  \\
Neuroticism &  &  &  &  &  &  &  &  &  & $0.032$ & $0.446$ &  \\
\multicolumn{13}{>{\color{editCol}}l}{\textit{Feedback Orientation Scale}} \\
Feedback Utility &  &  &  &  &  &  &  &  &  & $0.157$ & $0.035$ & * \\
Feedback Accountability &  &  &  &  &  &  &  &  &  & $-0.036$ & $0.678$ &  \\
Feedback Social Awareness &  &  &  &  &  &  &  &  &  & $0.085$ & $0.144$ &  \\
Feedback Self-Efficacy &  &  &  &  &  &  &  &  &  & $0.096$ & $0.131$ &  \\
\multicolumn{13}{>{\color{editCol}}l}{\textit{Social Anxiety (ref: High)}} \\
\quad Low &  &  &  &  &  &  &  &  &  & $-0.120$ & $0.174$ &  \\
\hline
\arrayrulecolor{black}
Conditional $R^2$ & \multicolumn{3}{>{\color{editCol}}c|}{0.447} & \multicolumn{3}{>{\color{editCol}}c|}{0.464} & \multicolumn{3}{>{\color{editCol}}c|}{0.469} & \multicolumn{3}{>{\color{editCol}}c}{0.478} \\
\hline
\end{tabular}
\label{apx_tab:rq2_full_hs}
\end{table*}

\definecolor{editCol}{RGB}{0,0,0}
\definecolor{lightline}{RGB}{210,210,210}
\begin{table*}[th!]
\centering
\scriptsize
\setlength{\tabcolsep}{6pt}
\caption{\textcolor{editCol}{\textbf{Incrementally built models for RQ2}: estimated effects from Linear Mixed-Effects Models for \dvcc{} ($M_{RQ2,CC}$). $M_{RQ2,CC,a}$ includes the most basic model. Intermediate models are provided, where $M_{RQ2,CC,b}$ adds demographics and work context; $M_{RQ2,CC,c}$ adds organizational context; and $M_{RQ2,CC}$ adds individual personality to form the full model. Table entries report coefficient estimates and corresponding $p$-values ($\cdot\ p < 0.10$, * $p < 0.05$, ** $p < 0.01$, *** $p < 0.001$).}}
\begin{tabular}{>{\color{editCol}}r|>{\color{editCol}}l>{\color{editCol}}l>{\color{editCol}}l|>{\color{editCol}}l>{\color{editCol}}l>{\color{editCol}}l|>{\color{editCol}}l>{\color{editCol}}l>{\color{editCol}}l|>{\color{editCol}}l>{\color{editCol}}l>{\color{editCol}}l}
& \multicolumn{3}{>{\color{editCol}}c|}{$M_{RQ2,CC,a}$} & \multicolumn{3}{>{\color{editCol}}c|}{$M_{RQ2,CC,b}$} & \multicolumn{3}{>{\color{editCol}}c|}{$M_{RQ2,CC,c}$} & \multicolumn{3}{>{\color{editCol}}c}{$M_{RQ2,CC}$} \\
& Est. & \multicolumn{2}{>{\color{editCol}}c|}{$p$-value} & Est. & \multicolumn{2}{>{\color{editCol}}c|}{$p$-value} & Est. & \multicolumn{2}{>{\color{editCol}}c|}{$p$-value} & Est. & \multicolumn{2}{>{\color{editCol}}c}{$p$-value} \\
\hline
\arrayrulecolor{lightline}
\multicolumn{13}{>{\color{editCol}}l}{\textbf{H2: \tasktype[Task Type] (ref: \assurance{})}} \\
\quad \brainstorming{} & $-0.285$ & $< .001$ & *** & $-0.286$ & $< .001$ & *** & $-0.286$ & $< .001$ & *** & $-0.286$ & $< .001$ & *** \\
\quad \proofing{} & $-0.237$ & $< .001$ & *** & $-0.237$ & $< .001$ & *** & $-0.237$ & $< .001$ & *** & $-0.237$ & $< .001$ & *** \\
\quad \search{} & $-0.331$ & $< .001$ & *** & $-0.331$ & $< .001$ & *** & $-0.331$ & $< .001$ & *** & $-0.332$ & $< .001$ & *** \\
\hline
\multicolumn{13}{>{\color{editCol}}l}{\textbf{H3: \role[Role] (ref: \assisting{})}} \\
\quad \delegating{} & $0.035$ & $0.288$ &  & $0.035$ & $0.288$ &  & $0.034$ & $0.289$ &  & $0.034$ & $0.290$ &  \\
\hline
\multicolumn{13}{>{\color{editCol}}l}{\textbf{H4: \specialization[Specialization] (ref: \openassist{})}} \\
\quad \corpassist{} & $0.006$ & $0.864$ &  & $0.006$ & $0.862$ &  & $0.006$ & $0.863$ &  & $0.006$ & $0.860$ &  \\
\hline
\multicolumn{13}{>{\color{editCol}}l}{\textbf{H5: \visibility[Visibility] (ref: \shared{})}} \\
\quad \personal{} & $-0.342$ & $< .001$ & *** & $-0.342$ & $< .001$ & *** & $-0.342$ & $< .001$ & *** & $-0.341$ & $< .001$ & *** \\
\hline
\multicolumn{13}{>{\color{editCol}}l}{\textbf{Vignette Order}} \\
Vignette Number & $-0.009$ & $0.153$ &  & $-0.009$ & $0.163$ &  & $-0.009$ & $0.155$ &  & $-0.009$ & $0.152$ &  \\
\hline
\multicolumn{13}{>{\color{editCol}}l}{\textbf{Demographics and Work}} \\
\multicolumn{13}{>{\color{editCol}}l}{\textit{Gender (ref: Man)}} \\
\quad Woman &  &  &  & $0.016$ & $0.850$ &  & $-0.016$ & $0.857$ &  & $0.005$ & $0.953$ &  \\
\quad Other &  &  &  & $0.124$ & $0.702$ &  & $0.029$ & $0.929$ &  & $-0.041$ & $0.901$ &  \\
Age (Ordinal Rank) &  &  &  & $-0.095$ & $0.035$ & * & $-0.094$ & $0.035$ & * & $-0.091$ & $0.051$ & $\cdot$ \\
\multicolumn{13}{>{\color{editCol}}l}{\textit{Job Function (ref: Analysis / Research)}} \\
\quad Software / Engineering &  &  &  & $-0.025$ & $0.831$ &  & $-0.057$ & $0.619$ &  & $-0.020$ & $0.870$ &  \\
\quad Planning / Project Management &  &  &  & $-0.198$ & $0.106$ &  & $-0.206$ & $0.094$ & $\cdot$ & $-0.168$ & $0.180$ &  \\
\quad Writing / Content Creation &  &  &  & $-0.030$ & $0.830$ &  & $0.026$ & $0.850$ &  & $0.060$ & $0.671$ &  \\
\quad Management / Strategy &  &  &  & $0.028$ & $0.867$ &  & $0.029$ & $0.862$ &  & $0.035$ & $0.840$ &  \\
\quad Design / Creative Work &  &  &  & $0.145$ & $0.394$ &  & $0.230$ & $0.195$ &  & $0.289$ & $0.109$ &  \\
\multicolumn{13}{>{\color{editCol}}l}{\textit{Job Role (ref: Managers)}} \\
\quad Skilled Office Worker &  &  &  & $-0.361$ & $0.029$ & * & $-0.371$ & $0.026$ & * & $-0.398$ & $0.019$ & * \\
\quad Senior Staff &  &  &  & $-0.351$ & $0.034$ & * & $-0.366$ & $0.029$ & * & $-0.395$ & $0.022$ & * \\
\multicolumn{13}{>{\color{editCol}}l}{\textit{Tenure (ref: $\leq$2 years)}} \\
\quad 3--5 years &  &  &  & $0.030$ & $0.761$ &  & $-0.001$ & $0.989$ &  & $0.020$ & $0.841$ &  \\
\quad 6+ years &  &  &  & $0.158$ & $0.160$ &  & $0.098$ & $0.388$ &  & $0.087$ & $0.457$ &  \\
\multicolumn{13}{>{\color{editCol}}l}{\textit{Work Arrangement (ref: In-Person)}} \\
\quad Remote &  &  &  & $0.015$ & $0.893$ &  & $-0.005$ & $0.961$ &  & $-0.006$ & $0.956$ &  \\
\quad Hybrid &  &  &  & $0.014$ & $0.896$ &  & $-0.018$ & $0.866$ &  & $0.000$ & $0.997$ &  \\
Has Direct Reports &  &  &  & $-0.101$ & $0.452$ &  & $-0.076$ & $0.573$ &  & $-0.091$ & $0.504$ &  \\
\hline
\multicolumn{13}{>{\color{editCol}}l}{\textbf{Organizational Context}} \\
Collaboration Frequency &  &  &  & $0.003$ & $0.962$ &  & $-0.003$ & $0.955$ &  & $-0.051$ & $0.423$ &  \\
Number of Colleagues &  &  &  & $0.036$ & $0.526$ &  & $-0.002$ & $0.972$ &  & $0.008$ & $0.899$ &  \\
Org. Has AI Training &  &  &  &  &  &  & $-0.032$ & $0.407$ &  & $-0.054$ & $0.184$ &  \\
Org. Encourages AI &  &  &  &  &  &  & $0.038$ & $0.498$ &  & $0.022$ & $0.697$ &  \\
\multicolumn{13}{>{\color{editCol}}l}{\textit{Organization Size (ref: <50)}} \\
\quad 50-249 &  &  &  &  &  &  & $-0.052$ & $0.746$ &  & $-0.093$ & $0.571$ &  \\
\quad 250-999 &  &  &  &  &  &  & $-0.101$ & $0.542$ &  & $-0.128$ & $0.455$ &  \\
\quad 1,000-9,999 &  &  &  &  &  &  & $0.085$ & $0.577$ &  & $0.088$ & $0.574$ &  \\
\quad 10,000 or more &  &  &  &  &  &  & $0.272$ & $0.097$ & $\cdot$ & $0.297$ & $0.075$ & $\cdot$ \\
\hline
\multicolumn{13}{>{\color{editCol}}l}{\textbf{Individual Personality}} \\
Team Psychological Safety &  &  &  &  &  &  &  &  &  & $0.032$ & $0.483$ &  \\
\multicolumn{13}{>{\color{editCol}}l}{\textit{Big-5 Personality}} \\
Openness &  &  &  &  &  &  &  &  &  & $0.014$ & $0.727$ &  \\
Conscientiousness &  &  &  &  &  &  &  &  &  & $0.030$ & $0.607$ &  \\
Extraversion &  &  &  &  &  &  &  &  &  & $-0.014$ & $0.737$ &  \\
Agreeableness &  &  &  &  &  &  &  &  &  & $0.033$ & $0.477$ &  \\
Neuroticism &  &  &  &  &  &  &  &  &  & $0.004$ & $0.935$ &  \\
\multicolumn{13}{>{\color{editCol}}l}{\textit{Feedback Orientation Scale}} \\
Feedback Utility &  &  &  &  &  &  &  &  &  & $0.124$ & $0.128$ &  \\
Feedback Accountability &  &  &  &  &  &  &  &  &  & $-0.062$ & $0.510$ &  \\
Feedback Social Awareness &  &  &  &  &  &  &  &  &  & $0.068$ & $0.283$ &  \\
Feedback Self-Efficacy &  &  &  &  &  &  &  &  &  & $0.077$ & $0.267$ &  \\
\multicolumn{13}{>{\color{editCol}}l}{\textit{Social Anxiety (ref: High)}} \\
\quad Low &  &  &  &  &  &  &  &  &  & $-0.031$ & $0.746$ &  \\
\hline
\arrayrulecolor{black}
Conditional $R^2$ & \multicolumn{3}{>{\color{editCol}}c|}{0.404} & \multicolumn{3}{>{\color{editCol}}c|}{0.421} & \multicolumn{3}{>{\color{editCol}}c|}{0.426} & \multicolumn{3}{>{\color{editCol}}c}{0.436} \\
\hline
\end{tabular}
\label{apx_tab:rq2_full_cc}
\end{table*}


\end{document}